\documentclass[preprint]{elsarticle}

\usepackage{times}
\usepackage{array}
\usepackage{graphicx} 
\usepackage{mathtools} 
\usepackage[version=3]{mhchem} 
\usepackage{amsmath,amssymb}
\usepackage{appendix} 
\usepackage{fixltx2e} 
\usepackage{gensymb} 
\usepackage{caption} 
\usepackage{subcaption} 
\usepackage{booktabs} 
\usepackage{multirow}

\newcommand{\ra}[1]{\renewcommand{\arraystretch}{#1}}

\biboptions{sort&compress, square, comma}
\usepackage[hidelinks]{hyperref}

\usepackage{url, breakurl}

\usepackage[english]{babel}
\usepackage{textcomp}

\usepackage{siunitx} 
\DeclareSIUnit\atm{atm}

\usepackage{soul}
\usepackage[usenames, dvipsnames]{color}

\usepackage{etoolbox}
\newbool{AltEngPlots}

\graphicspath{{./Figures/}}

\DeclarePairedDelimiter{\abs}{\lvert}{\rvert}
\DeclarePairedDelimiter{\set}{\lbrace}{\rbrace}
\DeclareMathSymbol{\mlq}{\mathord}{operators}{``}
\DeclareMathSymbol{\mrq}{\mathord}{operators}{`'}
\newcommand{\Edrgep}{\varepsilon_{\text{\tiny DRGEP}}}
\newcommand{\Edrg}{\varepsilon_{\text{\tiny DRG}}}
\newcommand{\liangtargets}{\textbraceleft\ce{n{\hyphen}C7H16}, \ce{CO}, \ce{HO2}\textbraceright}
\newcommand{\liangtargetsipent}{\textbraceleft\ce{i{\hyphen}C5H11OH}, \ce{CO}, \ce{HO2}\textbraceright}
\newcommand{\fulltargets}{\textbraceleft\ce{n{\hyphen}C7H16}, \ce{CO}, \ce{HO2}, \ce{OH}\textbraceright}
\newcommand{\fulltargetsipent}{\textbraceleft\ce{i{\hyphen}C5H11OH}, \ce{CO}, \ce{HO2}, \ce{OH}\textbraceright}
\newcommand{\DIC}[1]{$r_{#1}$}
\newcommand{\DICeq}[1]{r_{#1}}

\begin{document}

\title{An automated target species selection method for dynamic adaptive chemistry simulations}
\journal{Combustion and Flame}

\author[uconn]{Nicholas J.~Curtis}
\author[oregon]{Kyle E.~Niemeyer}
\author[uconn]{Chih-Jen Sung\corref{cor1}}
\ead{cjsung@engr.uconn.edu}

\address[uconn]{Department of Mechanical Engineering, University of Connecticut, Storrs, CT, 06269, USA}
\address[oregon]{School of Mechanical, Industrial, and Manufacturing Engineering, Oregon State University, Corvallis, OR 97331, USA\vspace{-4ex}}
\date{\today}

\cortext[cor1]{Corresponding author}

\begin{abstract}
The relative importance index (RII) method for determining appropriate target species for dynamic adaptive chemistry (DAC) simulations using the directed relation graph with error propagation (DRGEP) method is developed.  The adequacy and effectiveness of this RII method is validated for two fuels: n-heptane and isopentanol, representatives of a ground transportation fuel component and bio-alcohol, respectively.

The conventional method of DRGEP target species selection involves picking an unchanging (static) set of target species based on the combustion processes of interest; however, these static target species may not remain important throughout the entire combustion simulation, adversely affecting the accuracy of the method.
In particular, this behavior may significantly reduce the accuracy of the DRGEP-based DAC approach in complex multidimensional simulations where the encountered combustion conditions cannot be known a priori with high certainty.
Moreover, testing multiple sets of static target species to ensure the accuracy of the method is generally computationally prohibitive.
Instead, the RII method determines appropriate DRGEP target species solely from the local thermo-chemical state of the simulation, ensuring that accuracy will be maintained.
Further, the RII method reduces the expertise required of users to select DRGEP target species sets appropriate to the combustion phenomena under consideration. 

Constant volume autoignition simulations run over a wide range of initial conditions using detailed reaction mechanisms for n-heptane and isopentanol show that the RII method is able to maintain accuracy even when traditional static target species sets fail, and are even more accurate than expert-selected target species sets.
Additionally, the accuracy and efficiency of the RII method are compared to those of static target species sets in single-cell engine simulations under homogeneous charge compression ignition conditions.  
For simulations using more stringent DRGEP thresholds, the RII method performs similarly to that of the static target species sets.
With a larger DRGEP threshold, the RII method is significantly more accurate than the static target species sets without imposing significant computational overhead.

Furthermore, the applicability of the RII method to a DRG-based DAC scheme is discussed.
\end{abstract}

\maketitle

\section{Introduction}

The use of detailed reaction mechanisms is essential for high-fidelity predictions of important combustion phenomena such as pollutant emissions~\cite{correa_turbines} and local flame extinction~\cite{lu_towardrealistic}---a cause of lean blowout \cite{shan_leanblowout}---as well as in the design of next-generation combustion devices such as homogeneous charge compression ignition (HCCI) engines \cite{yao_hcci}.
However, the large size and high chemical stiffness of transportation-relevant fuel mechanisms prohibit their use in realistic simulations.
For multidimensional reacting flow simulations, the chemistry time integration can take up to \SIrange{75}{99}{\percent} of the total simulation time \cite{tonse_cfd,liang_semiimplicit,shi_hybrid,shi_dac}.
In order to utilize large chemical reaction mechanisms for transportation-relevant fuels in realistic simulations, accurate mechanism reduction and chemical stiffness removal strategies must be employed.

Detailed reaction mechanisms are constructed to be valid over a wide range of thermo-chemical states, and therefore tend to contain many species and reactions that are not important in all combustion regimes.
A skeletal mechanism is created by removing species and reactions considered unimportant for the thermo-chemical state space under consideration.
A comprehensive skeletal mechanism is constructed for a broad thermo-chemical state space, while a local skeletal mechanism is created for a specific, limited range of thermo-chemical states.

Several systematic techniques to generate skeletal mechanisms by removal of unimportant species and reactions from a detailed mechanism have been developed; the directed relation graph (DRG) method \cite{drg_lu,drg_lu2,stategies_lu} and the directed relation graph with error propagation (DRGEP) \cite{drgep_pepiot} are two skeletal reduction methods popularly used due to their efficiency and reliability.
Other commonly used skeletal reduction methods include sensitivity analysis~\cite{rabitz_1983,turanyi_sa_1,turanyi_sa_2}, principal component analysis~\cite{vajda_pca}, level of importance analysis~\cite{lovas_loi,lovas_loi2,lovas_2009}, and methods based on computational singular perturbation~\cite{csp1,csp2,csp3} modified for skeletal reduction \cite{valorani_csp,valorani_csp2}.
To generate even more compact skeletal mechanisms, the DRG and DRGEP methods are often combined with sensitivity analysis, as in the DRG-aided sensitivity analysis~\cite{stategies_lu} and DRGEP-aided sensitivity analysis~\cite{niemeyer_drgepsa, Niemeyer:2014} methods.

In order to further optimize comprehensive skeletal mechanisms, dimension reduction methods are often employed. For example, chemical timescale analysis is utilized to exploit the tendency of reaction mechanisms to be attracted to lower dimensional manifolds, constraining the dimensionality of the reaction mechanism and thus simplifying integration.
Methods such as the quasi-steady state~\cite{qssa} and partial equilibrium approximations~\cite{pe_approx1,pe_approx2} assume a species or reaction quickly reaches a steady state after initial transience, and thus can be solved for algebraically.
More systematic dimension reduction methods include the computational singular perturbation~\cite{csp1,csp2,csp3} and the intrinsic low-dimensional manifold~\cite{ildm} methods, which analyze the Jacobian matrix to decouple the fast and slow chemical reaction modes to reduce chemical stiffness.

Finally, to accelerate chemical integration, tabulation methods that store and reuse previous solution information to cheaply construct approximate solutions of chemical integrations are often used.
Common examples include in situ adaptive tabulation~\cite{pope1997computationally} (ISAT), piecewise reusable implementation of solution mapping~\cite{prism}, and artificial neural networks~\cite{Christo1996}.

Most skeletal reduction approaches create a single, comprehensive skeletal mechanism for use over a prescribed range of conditions expected to be encountered.
However, this approach is inherently at a disadvantage in multidimensional simulations because the same level of detail must be applied to the entire domain.
Inside the flame zone, a highly detailed skeletal mechanism will likely be necessary to maintain accuracy, but in computational cells where combustion is mostly completed or weakly reacting a much smaller skeletal mechanism may be sufficient.

Recently, Liang et al.~\cite{liang_dac,liang_dac_gas} proposed a dynamic adaptive chemistry (DAC) scheme to exploit this observation, where the DRGEP skeletal reduction method was applied to a thermo-chemical state under consideration to generate a smaller, locally accurate skeletal mechanism.
The local skeletal mechanism was then integrated for a single simulation time-step and discarded; this process was repeated at the next time step.
The reduced expense resulting from integrating the smaller, locally accurate mechanism outweighed the overhead of the reduction method, leading to time savings overall.
For this reason, reduction methods that scale linearly with the problem size (e.g.,\ DRG, DRGEP, element flux analysis~\cite{he_element}) are typically used in DAC schemes.

Liang et al.~\cite{liang_dac} utilized this method to achieve a \num{30}-fold speedup with high accuracy for single cell HCCI simulations of n-heptane.
Later, they demonstrated the applicability of the method for n-heptane\slash isooctane\slash toluene blends in HCCI and homogeneous autoignition simulations~\cite{liang_dac_gas}.
By pairing an element flux analysis method with the DAC scheme, He et al.~\cite{he_element} achieved a \num{25}-fold speedup in a simulation of n-pentane in a pairwise mixed stirred reactor; however, the  overhead of the flux-based reduction method consumed nearly \SI{20}{\percent} of the total simulation time.
Yang et al.~\cite{yang_dac_drg} paired the DAC scheme with the DRG method in turbulent methane flame simulations.
They found that the DRG-based DAC method accurately reproduced the combustion process of a partially stirred reactor with significant levels of non-equilibrium chemistry.
Furthermore, for simulations with longer flow time scales (e.g., \SI{e-4}{\second} or longer) it was more efficient to generate a larger skeletal mechanism to be used for the whole flow time step, rather than performing multiple reduction\slash integrations steps within a single flow time step due to the overhead of the mechanism reduction \cite{yang_dac_drg}.

Tosatto et al.~\cite{tosatto_drg} formulated a DRG-based DAC scheme that additionally considered transport fluxes, achieving speedup factors of 5 and 10 for a steady JP-8 flame and a time-dependent ethylene flame, respectively.
Gou et al.~\cite{dac_gou} paired a simplified version of the path flux analysis method~\cite{sun_2010} with the DAC scheme to develop a method of error control for DAC simulations.
Data tabulated from simple zero-dimensional simulations were combined with a reaction progress variable---the mass fraction of oxygen---to automatically select appropriate reduction thresholds during the simulation.
This method reached speedup factors of \numrange[range-phrase = --]{5}{100} with high accuracy.
However, the use of pre-tabulated data and choice of reaction progress variable may lack generality in turbulent reacting systems where the combustion conditions are unknown a priori and mixing plays a much stronger role.
Therefore, more investigation is needed in this direction.

Contino et al. \cite{Contino2011, contino2012simulations} proposed the tabulation of dynamic adaptive chemistry (TDAC) method that combines the strengths of tabulation methods (e.g., ISAT) and the DAC scheme. 
In the base ISAT method, when integrating the reaction mechanism at a thermo-chemical state, a database is first queried to determine if a similar state (and corresponding state after integration) are stored.
If an appropriate state is found, an approximation to the integration of the queried state is cheaply constructed from the stored data.
Otherwise, the reaction mechanism is directly integrated at the queried state and is used along with the resulting state to update the database.
In the TDAC method, the direct integration of the reaction mechanism is accelerated using a DAC scheme resulting in further computational savings.
A speedup factor of $\sim$500 was reported for premixed combustion cases, and $\sim$9 for non-premixed cases with high fidelity predictions of species concentrations and pressure traces.
Ren et al. \cite{ren_2014, ren_2014b} applied a similar scheme to computationally intensive partially stirred reactor simulations, showing up to $\sim$1000 speedup factor for premixed cases and a \SI{30}{\percent} improvement in computational efficiency over the base ISAT method for a non-premixed case. 

Apart from the reduction thresholds explored by Gou et al.~\cite{dac_gou}, the other major factor controlling the performance and accuracy of DRG\slash DRGEP-based DAC methods is the selection of target species (i.e.,\ search-initiating species).
These species are selected for their expected importance to the combustion processes under consideration, and control the reduction of the detailed mechanism.  The skeletal mechanisms generated by DRG\slash DRGEP-based methods include only species whose removal would introduce large error into the production or consumption of these target species.
Traditionally, these species are selected before the simulation and treated as target species throughout; however, as discussed by Shi et al.~\cite{shi_dac}, this methodology may overestimate the importance of these target species in certain combustion regimes.
For example, while the fuel molecule is almost always used as a target species, in high-temperature, post-ignition combustion almost all the large hydrocarbons have broken down into small molecules and the fuel molecule no longer plays an important role.
To address this problem, Shi et al.~\cite{shi_dac} proposed an extended DAC scheme (EDAC) that switches between a small number of target species sets based on the local thermo-chemical state, resulting in an additional \SIrange[range-phrase = --, range-units = single]{8}{10}{\percent} time savings in a three-dimensional direct-injection engine study using skeletal mechanisms for n-heptane.
As the EDAC method is limited to a handful of target species sets, it may be inaccurate if unanticipated combustion conditions are encountered.
Furthermore, it is not clear whether the EDAC method will easily extend to all problem types and reaction mechanisms, as much of the methodology for switching between the target species sets is empirically derived from zero-dimensional studies or taken from user experience.

This work will describe an automated target species selection method for dynamic adaptive chemistry simulations using only the local thermo-chemical state.
First, the DAC scheme and DRGEP method will be outlined in more detail in Section~\ref{sec:DAC Section}, before examining current target species selection methods for the DRGEP method and further demonstrating the need for dynamic target selection.
The relative importance index (RII) method of DRGEP target species selection will then be developed in Section~\ref{sec:RII}.
In Section~\ref{Validation}, the RII method will be validated and its performance will be compared to the current method of static target species selection.
Conclusions and suggestions for future directions will be presented in Section~\ref{sec:Conclusion}.

\section{Dynamic Adaptive Chemistry}
\label{sec:DAC Section}

\subsection{Directed Relation Graph with Error Propagation}
\label{sec:DRGEP}

The DRGEP method as proposed by Pepiot-Desjardins and Pitsch~\cite{drgep_pepiot} determines important species to be kept in the resulting skeletal mechanism based on their coupling to a list of target species (i.e.,\ search-initiating species) that are expected to be key to the combustion processes under consideration.
Each species in the detailed mechanism is represented as a vertex on a graph, and edges between species are defined using a direct interaction coefficient (DIC) that quantifies species coupling based on species production and consumption rates:

\begin{equation}
\DICeq{AB} = \frac{\abs*{\sum_{i = 1}^{N_R} \nu_{A,i} \omega_i \delta_{B,i} }}{\max (P_A, C_A)} \;,
\label{eq:DIC}
\end{equation}
where $P_A$, $C_A$, and $\delta_{B,i}$ are defined as:
\begin{equation}
P_A = \sum_{i = 1}^{N_R} \max(0, \nu_{A,i}\omega_i) \;,
\end{equation}
\begin{equation}
C_A = \sum_{i = 1}^{N_R} \max(0, -\nu_{A,i}\omega_i) \;,
\end{equation}
\begin{equation}
\delta_{B,i} = 
\begin{cases}
1& \text{if reaction $i$ involves species $B$} \;, \\
0& \text{otherwise} \;, \\
\end{cases}
\end{equation}
$\nu_{A,i}$ is the net stoichiometric coefficient of species $A$ in reaction $i$, $\omega_i$ is the net reaction rate of reaction $i$, and $N_R$ is the number of reactions in the mechanism.
The DIC measures the importance of species $B$ to species $A$, and represents the error in the overall production of $A$ that would result if $B$ was removed from the mechanism.
This definition of the DIC is modified from that of the original DRG method~\cite{drg_lu} in order to address shortcomings in situations with long chemical paths involving fast chemical modes~\cite{drgep_pepiot}. 

The DRGEP method also considers the propagation of error caused by removal of a species along reaction pathways; a path dependent interaction coefficient from a target species $T_j$ to a species $B$ along a pathway $p$ is defined as:
\begin{equation}
\DICeq{T_jB, p} = \prod\limits_{i=1}^{length(p) - 1} \DICeq{S_iS_{i+1}} \;,
\end{equation}
where the $i$th edge of path $p$ connects species $S_i$ and $S_{i + 1}$. The interaction coefficient between species $B$ and target $T_j$ is then defined as
\begin{equation}
R_{B,T_{j}} = \max_{\text{all paths } p}(\DICeq{T_jB, p}) \;.
\end{equation}
Finally, the overall interaction coefficient (OIC) is:
\begin{equation}
R_{B} = \max_{T_j \in \set{Targets}} R_{B,T_{j}} \;.
\end{equation}
The species $B$ is then considered active in the resultant skeletal mechanism if and only if
\begin{equation}
R_B \geq \Edrgep \;,
\end{equation}
where $\Edrgep$ is a specified DRGEP threshold value (e.g.,\ \num{e-4}).

Using this definition of the OIC, the error induced by removal of a species $B$ must propagate along the graph pathways to reach the target species set; a species further away on the graph from the target set is more likely to be removed for this reason.
In contrast, the DRG method uses a threshold $\Edrg$ to determine \textit{whether} an edge exists between two species (the edge exists only if the DRG DIC is greater than $\Edrg$), and a species $B$ is kept in the skeletal mechanism if there exists a path from the target set to $B$.
This binary truncation of the DIC eliminates valuable information on species couplings, and the DRG method will generally produce larger skeletal mechanisms than the DRGEP method as a result~\cite{drgep_pepiot,niemeyer_drgepsa}.  

Although the error propagation step of the DRGEP method makes a greater reduction extent possible, the tendency to remove species further away from the targets makes the accuracy of the DRGEP method highly dependent on the proper selection of target species.
In Section~\ref{sec:Applicability} it is demonstrated that large errors can rapidly accumulate when improper target species are selected.
In contrast, when using the DRG method any species with edges connecting to a target species will be included in the skeletal mechanism, regardless of their proximity to the target species, and the accuracy of the resulting skeletal mechanism will be less dependent on target species selection.

The direct interaction coefficients are calculated using the linear-time calculation approach (looping over reactions) proposed by Lu and Law \cite{lu_2006}.
Following the work of Niemeyer and Sung \cite{niemeyer_graphsearch}, Dijkstra's algorithm (implemented with a binary heap) is used to calculate the OICs.
Additionally, following ideas from the RBFS algorithm presented by Liang et al.~\cite{liang_dac}, edges smaller than the DRGEP threshold are not expanded during the graph search, and the search exits when the maximum value on the heap is less than the DRGEP threshold.

\subsection{Dynamic Adaptive Chemistry Scheme Formulation}
In the dynamic adaptive chemistry scheme, the starting mechanism is first reduced (e.g.,\ using the DRGEP method), resulting in a skeletal mechanism with $m$ active species (superscript $a$) and $n$ inactive species (superscript $i$).
For a fixed mass, known volume system, the chemical kinetics equations can be expressed as:
\begin{equation}
\begin{pmatrix}
\dot{y}^a_1 \\ \vdots \\ \dot{y}^a_m \\ \dot{T}
\end{pmatrix}
 = 
\begin{pmatrix}
f_1(T, p, y^a_1, \ldots, y^a_m, y^i_1, \ldots, y^i_n) \\
\vdots \\
f_m(T, p, y^a_1, \ldots, y^a_m, y^i_1, \ldots, y^i_n) \\
f_{m+1}(T, p, y^a_1, \ldots, y^a_m, y^i_1, \ldots, y^i_n)
\end{pmatrix} \;,
\label{dac_eq}
\end{equation}
where $T$ represents the temperature, $p$ the pressure, $y^a_j$ the mass fraction of the $j$th of $m$ active species, and $y^i_k$ the mass fraction of the $k$th of $n$ inactive species.
A reaction is considered active (and thus considered in the right-hand-side functions $f_i$) if and only if all participating species are active in the skeletal mechanism.
It is noted that although inactive species do not participate directly in any active reaction, their removal can induce serious errors in third-body and pressure-dependent reactions.
In order to minimize the size of the ODE system while accounting for these third body effects, the DAC scheme calculates the net species production rates of the active species, as well as the derivatives of any state variables (e.g.,\ temperature, pressure) from the local thermo-chemical state---including inactive species---as seen in Eq.~\eqref{dac_eq}.

The skeletal mechanism is then integrated for one simulation time-step, holding the inactive species fixed, and the resulting composition is stored.
As the simulation time-step for a typical computational fluid dynamics (CFD) simulation is small (e.g.,\ \SI{e-5}{\second}) the local skeletal mechanism can be assumed to be valid for the whole time-step.  In some situations the flow time-step may be significantly larger, invalidating this assumption.  In such cases, multiple reduction steps within a single flow time-step are needed~\cite{yang_dac_drg}.

\subsection{Dynamic Adaptive Chemistry Implementation}

The DAC scheme used in this work follows that of Liang et al.~\cite{liang_dac}, as seen in Eq.~\eqref{dac_eq}.
All calculations are completed based on the open-source chemical kinetics software Cantera \cite{cantera2a11}, modified to allow dynamic mechanism reduction as well as to enable dynamic adaptive chemistry integration.
The simulation time-step is set to \SI{5e-6}{\second} for all cases, and the integrator (including Jacobian) is reinitialized at every time-step in order to account for the changing problem size.
This reinitialization adversely affects the performance of homogeneous simulations because the Jacobian must be recalculated at the beginning of each simulation time-step; however, it is realistic for operator-splitting\slash fractional-step schemes commonly used in reacting flow simulations where transport invalidates such saved information.
Engine simulations are run using constant time steps.  
At each time step, the crank angle and piston velocity are calculated using the following engine parameters taken from Sj\"{o}berg et al.~\cite{sjoberg_2007}: a displacement volume of \SI{0.981}{\liter}, a ratio of connecting rod length to crank radius of 3.2, and a compression ratio of 14.
For constant-volume autoignition runs, the ignition delay is determined as the time where temperature reached \SI{400}{\kelvin} greater than the initial temperature.
Simulation wall-clock times are reported as the average over 25 runs.
All simulations were performed on a single core of a 12-core Xeon X5650 processor.

\subsection{Conventional Target Species Selection Methods}

In order to investigate the performance of static target species sets, version 2 of the LLNL detailed n-heptane mechanism with 561 species and 4564 reactions~\cite{nhept_1,nhept_2,nhept_website} was used; n-heptane is an important primary reference fuel (PRF) for gasoline that has been extensively studied in the literature.
n-Heptane exhibits strong negative temperature coefficient (NTC) behavior, as well as two-stage ignition processes in the low temperature chemistry regime \cite{peters_ignition}.
These factors require a high-fidelity DAC scheme, as large errors can accumulate rapidly otherwise.
As such, the n-heptane mechanism is a suitable choice to demonstrate the problems with static target species selection.

The performance and accuracy of the DAC scheme are dependent on the reduction method utilized; for the DRGEP method, these are functions of the target species chosen and the DRGEP threshold ($\Edrgep$) used.
Due to the error propagation step, the DRGEP method is more sensitive to the proper selection of target species than the original DRG method.
Traditionally, target species have been chosen based on their expected importance to important combustion processes.
Typical choices of target species include the fuel, oxygen, combustion products (e.g.,\ \ce{CO2}), and certain key radicals and intermediates known to be good indicator species (e.g.,\ \ce{H}, \ce{OH}, \ce{CO}, \ce{HO2})~\cite{liang_dac_gas,yang_dac_drg,tosatto_drg}.
In addition, species such as \ce{NO_x} and other pollutants may be added to the target species if high accuracy in emission predictions are required.

A target species set that is not modified through addition or removal of target species throughout a simulation will be termed ``static'' (unchanging) in this context.
One common practice is to remove a species from the target set when its mass fraction falls below some small cutoff (e.g, \num{e-30}~\cite{liang_dac}).
In this work, the static target set category includes cases where the only changes are removals of target species from the set based on mass fraction criteria.

Most attempts to determine appropriate target species sets have consisted of directly comparing the accuracy and performance of a small number of static target species sets.
However, it is often difficult to choose a single static target species set that will be appropriate for all combustion processes.
As demonstrated by Shi et al.~\cite{shi_dac}, the choice of static target species can overestimate the importance of the targets for certain combustion stages.
For example, during the post-ignition period most of the large hydrocarbons have been broken down into small molecules, and thus the fuel species should no longer be considered as a target species.
The subsequent section will further detail the problems with static target species sets.

\subsection{Performance and Range of Applicability of Static Target Species Sets}
\label{sec:Applicability}

In a multidimensional combustion problem many different combustion modes will be encountered.
For example, these include local ignition and extinction, as well as a variety of temperatures, pressures, fuel blends, and local equivalence ratios~\cite{lu_towardrealistic}.
In order for a DAC scheme to produce high-fidelity results in such a simulation, it should remain accurate at all the possible combustion conditions that may be encountered.
In their original DAC study, Liang et al.~\cite{liang_dac} compared the performance of several static target species sets in single cell, adiabatic, HCCI engine simulations of n-heptane (\ce{n{\hyphen}C7H16}).
It was found that the target species set \liangtargets\ was capable of accurately reproducing both the pressure and major species traces with $\Edrgep = $ \num{e-4}, and mass fraction cutoff of \num{e-30}.

\begin{figure}
\begin{subcaptionbox}{Constant volume ignition delays at \SI{20}{\atm}\label{subfig:conv static igdelays}}[\linewidth]{
	\centering
	\includegraphics[width = \linewidth, keepaspectratio = true]{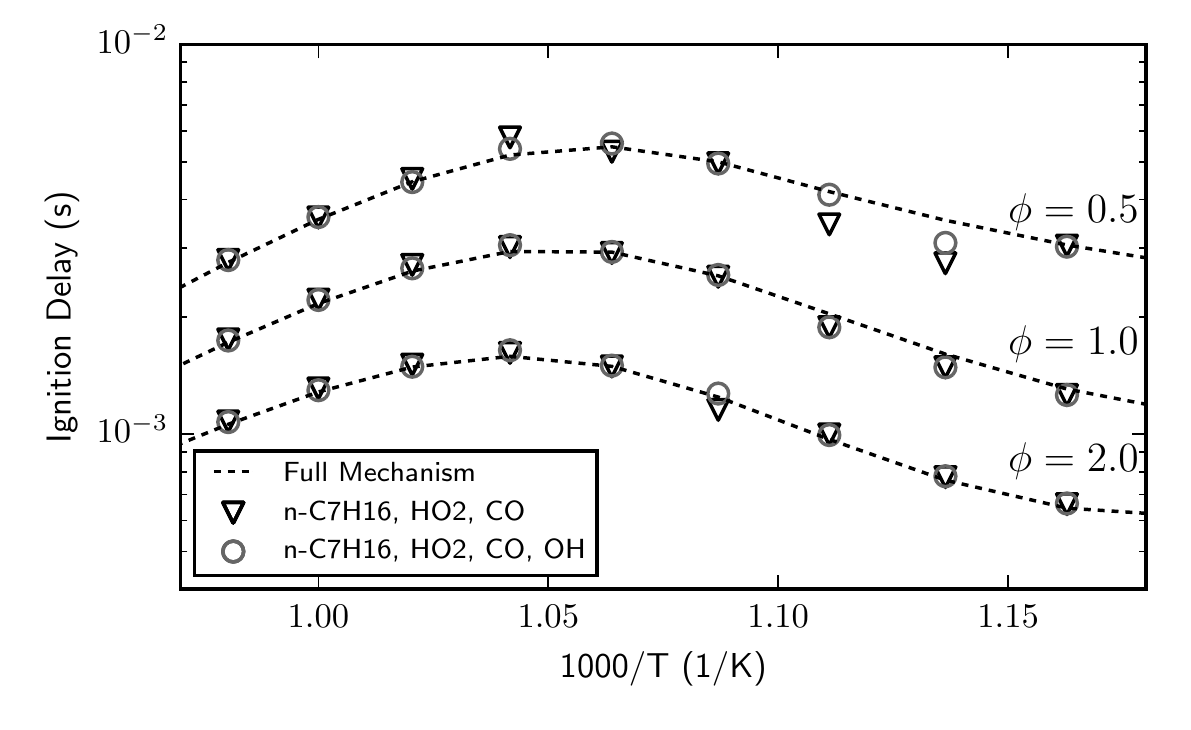}
}
\end{subcaptionbox}
\begin{subcaptionbox}{Temperature trace of constant-volume autoignition simulation at \SI{900}{\kelvin}, \SI{20}{\atm}, $\phi = 0.5$\label{subfig:conv igdelay temp trace}}[\linewidth]{
	\centering
	\includegraphics[width = \linewidth, keepaspectratio = true]{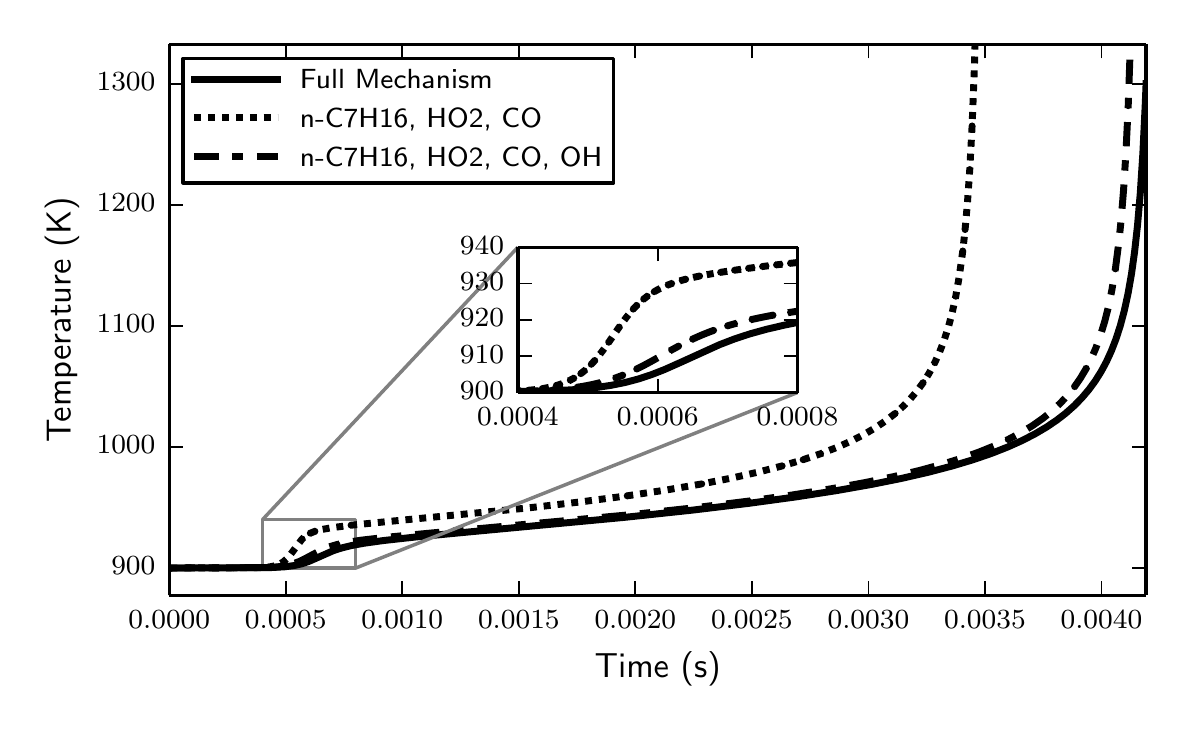}
}
\end{subcaptionbox}
\caption{In \subref{subfig:conv static igdelays} the constant-volume ignition delays using two static target species sets for n-heptane at \SI{20}{\atm}, with $\Edrgep = $ \num{e-4} and a mass fraction cutoff of \num{e-30} are displayed.  The static target species set without \ce{OH} has a maximum percent error of \SI{22}{\percent} while the set including \ce{OH} has a maximum error of just \SI{12.5}{\percent}.  In \subref{subfig:conv igdelay temp trace} the temperature trace of the two target sets at \SI{900}{\kelvin}, \SI{20}{\atm}, $\phi = 0.5$ is compared to that of the full mechanism; by adding \ce{OH} to the target species set, the percent error in ignition delay is decreased from \SI{17}{\percent} to just \SI{1.8}{\percent}.}
\label{fig:conv static err}
\end{figure}

Figure~\ref{fig:conv static err} compares the performance of the target species set \liangtargets\ to that of \fulltargets\ for constant-volume autoignition simulations of n-heptane at \SI{20}{\atm}, in the temperature range of \SIrange[range-phrase=--]{860}{1000}{\kelvin}, and equivalence ratios of $\phi = \set{0.5, 1, 2}$.
As seen in Fig.~\ref{subfig:conv static igdelays}, adding \ce{OH} to the target species set reduces the maximum percent error in ignition delay from \SI{22}{\percent} to \SI{12.5}{\percent}.
Figure~\ref{subfig:conv igdelay temp trace} compares the temperature traces of the two static target species sets  for \SI{900}{\kelvin}, \SI{20}{\atm}, and $\phi = 0.5$ with that of the full mechanism; the target set \liangtargets\ greatly underpredicts the ignition delay for first stage ignition, and the resulting combustion process differs significantly from that of the full mechanism.
Adding \ce{OH} to the target species set results in proper prediction of first-stage ignition and a reduction of the overall error in ignition delay prediction.
As demonstrated here, adding \ce{OH} to the target species set can be necessary to accurately predict low-temperature ignition; in a multidimensional simulation where local ignition events may be encountered, this could be vital to the accuracy of the simulation.

\begin{figure}
\centering
\begin{subcaptionbox}{Ignition delay error in crank angle degrees \label{subfig:igdelay_hcci_nhept_static}}[0.5\linewidth]{
	\centering
	\includegraphics[width = 0.5\linewidth, keepaspectratio = true]{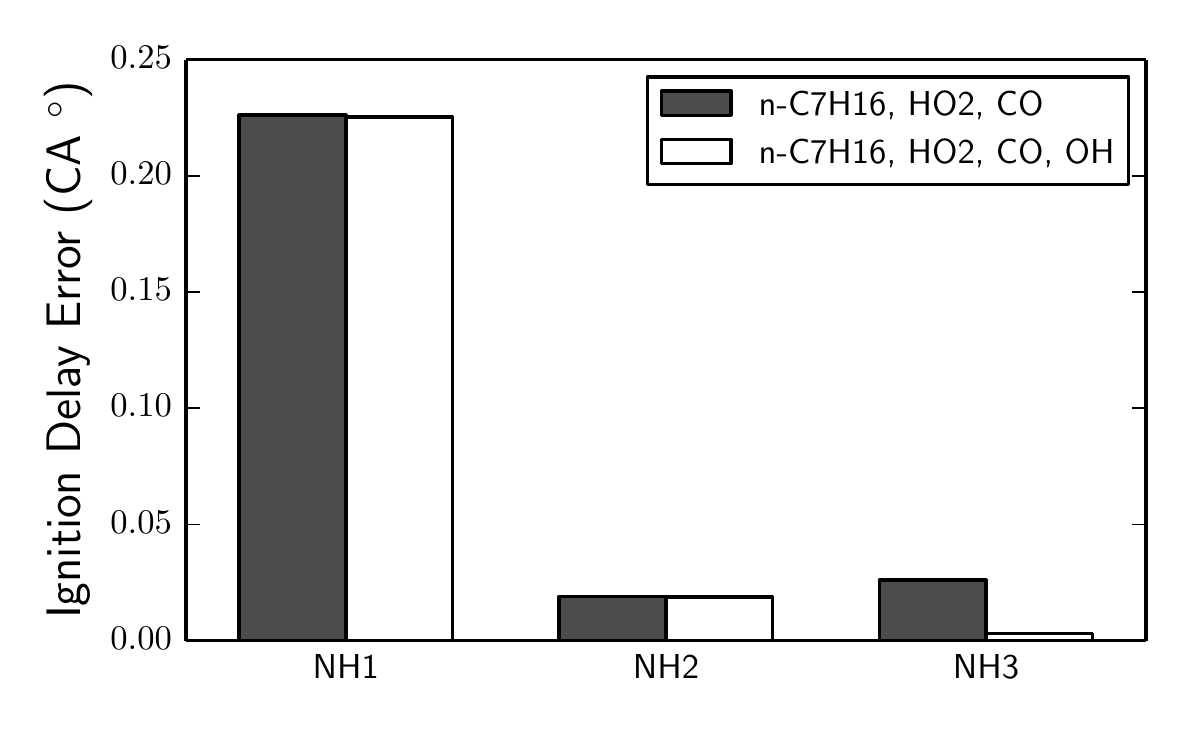}
}
\end{subcaptionbox} \\
\begin{subcaptionbox}{Normalized simulation wall time\label{subfig:walltime_hcci_nhept_static}}[0.5\linewidth]{
	\centering
	\includegraphics[width = 0.5\linewidth, keepaspectratio = true]{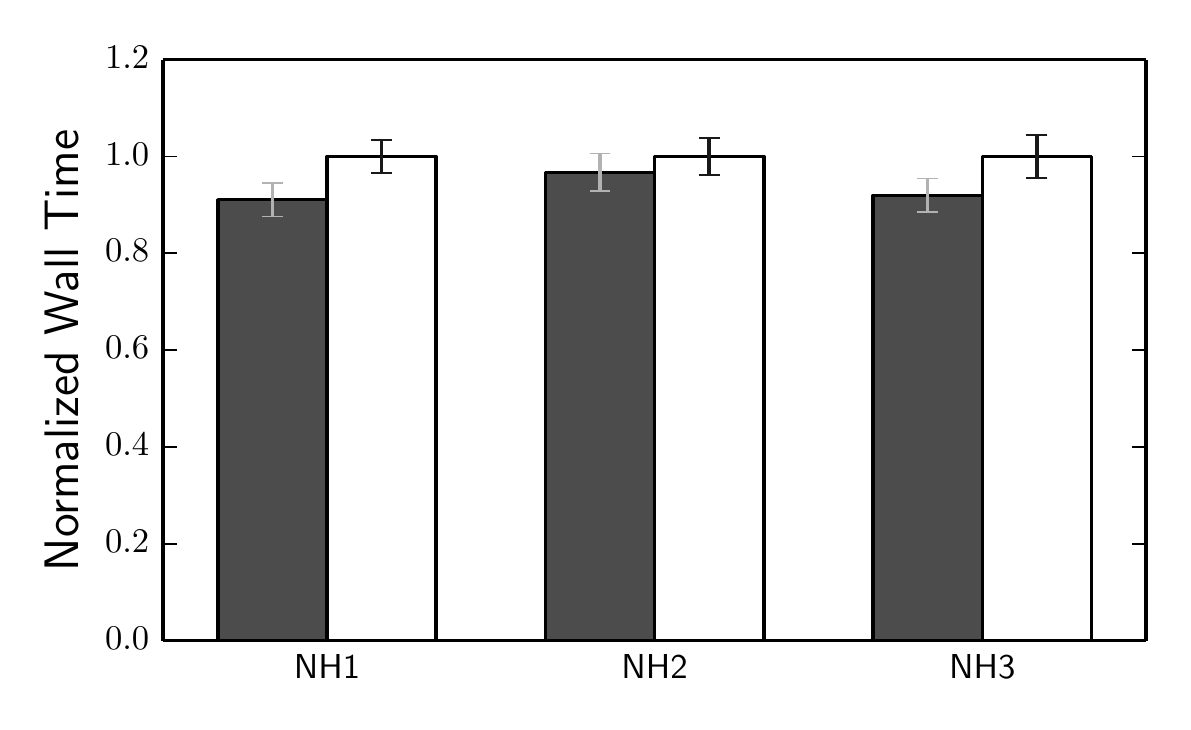}
}
\end{subcaptionbox}
\caption{Comparison of \subref{subfig:igdelay_hcci_nhept_static} ignition delay error and \subref{subfig:walltime_hcci_nhept_static} simulation wall time normalized by the longest wall time in each case for single-cell n-heptane engine simulations at the HCCI conditions listed in Table~\ref{tab:nhept_hcci_conditions}.  The two static target species sets were used with $\Edrgep = $ \num{e-4} and a mass fraction cutoff of \num{e-30}.}
\label{fig:nhept_static_hcci}
\end{figure}

However, the use of the target set \fulltargets\ is not without cost.
Figure~\ref{fig:nhept_static_hcci} compares the performance of the two static target sets at HCCI conditions adapted from Liang et al.~\cite{liang_dac}, using the engine parameters from Sj\"{o}berg et al.~\cite{sjoberg_2007} described above.
Both target sets reproduced the ignition delay predictions (in terms of crank angle degree) of the full detailed mechanism with high accuracy, resulting a maximum ignition delay error of \SI{0.23}{\degree}.
However, examining the simulation wall times shows that the target set \fulltargets\ is \SIrange[range-phrase = --, range-units=single]{5}{10}{\percent} slower than \liangtargets{}.
The use of a static target species set with suboptimal execution speed may be necessary in order to ensure that a simulation remains accurate for low-temperature ignition.

\begin{figure}
\centering
\begin{subfigure}[t]{\linewidth}
	\centering
	\includegraphics[width = \linewidth, keepaspectratio = true]{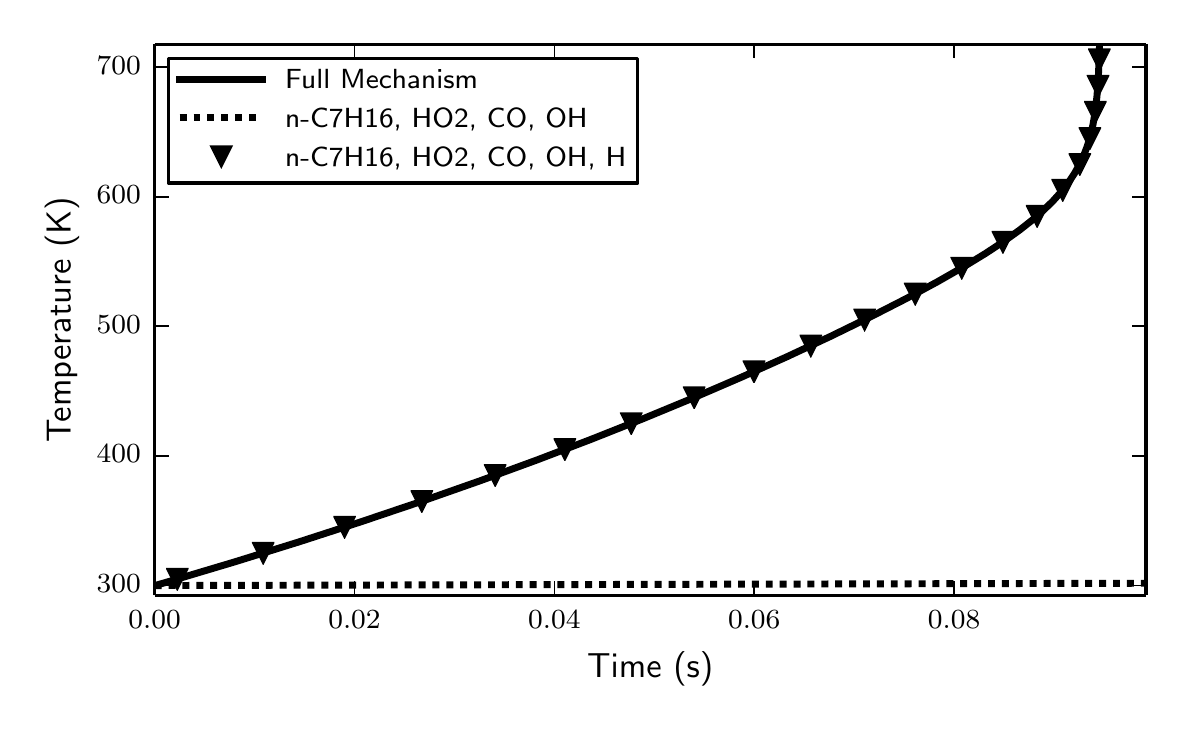}
\end{subfigure}
\caption{Temperature trace of two static target species sets for the constant-pressure reactor ignition of a stoichiometric n-heptane/air mixture by a pilot stream of the \ce{H} radical.  The mass flow rate of the pilot stream is 1:100 with that of the n-heptane/air mixture.}
\label{fig:diffusion-static}
\end{figure}

Furthermore, this is just one case where the static target species set \liangtargets\ may fail.
Consider for example that in diffusive systems, important highly mobile species such as the \ce{H} radical may need to be added to the target species set.
This is demonstrated in Fig.~\ref{fig:diffusion-static}, where a stream of stoichiometric n-heptane/air mixture at \SI{300}{\kelvin}, \SI{1}{\atm} flows into a constant-pressure reactor and is ignited by a pilot stream of the \ce{H} radical at the same temperature.
The static target species set \fulltargets\ is not sufficient to capture this ignition process; instead, the \ce{H} radical must be added to the target species set in order to maintain accuracy.
Although a special situation, this example demonstrates the inability of a static target species set to adapt to combustion conditions dominated by a species not included in the target set.
In order to ensure accuracy is maintained in a multidimensional simulation, other species corresponding to major combustion or physical processes may need to be added to the target set.
For static target species sets, this requires the user to determine which species may be needed, potentially adversely affecting performance if the proper selection is not made.

These issues with static target species selection motivate us to develop an automated target species selection method which can adapt to changing combustion conditions, ensuring that accuracy and execution speed are maintained. In due course, we will demonstrate that the developed method overcomes these problems and provides a promising alternative to static target species selection.

Finally, it is noted that although the accuracy of the DRG method is less sensitive to the selection of target species (as discussed in Section~\ref{sec:DRGEP}), DRG-based DAC schemes will also benefit from use of an automated target species selection method.
For example, as discussed previously the inclusion of the fuel molecule in the target species set during post-ignition may cause the resulting local skeletal mechanisms to be larger than necessary.  
Further, for cases where combustion is controlled by a species not included in the target species set (e.g. analogous to the example in Fig.~\ref{fig:diffusion-static}), large errors will still accumulate without manual determination of important species for the target species set.
The application of the developed automated target species selection method to DRG-based DAC schemes will address these issues, and is a topic that merits future investigation.

\subsection{Dynamic Target Species Selection}

The above observations suggest that no single static target species set will be optimal, in line with the work of Shi et al.~\cite{shi_dac} who proposed the EDAC scheme as a solution to this problem.
The EDAC scheme changes the target species set based on the local thermo-chemical state in order to better respond to changing combustion conditions and maintain computational efficiency.
However, the EDAC scheme uses empirically derived cutoffs of two proposed progress equivalence ratios as well as the local temperature to switch between various target species sets.
As a result, it may be difficult to extend to other combustion problems and reaction mechanisms.
Additionally, the EDAC scheme utilizes only a small fixed number of target species sets; the accuracy of the EDAC scheme will suffer if combustion conditions are encountered where none of the target species sets are appropriate.

As such, the goal of this work is to develop an automated method of determining appropriate target species solely from the local thermo-chemical state with the following characteristics:
\begin{enumerate}
\item The method will not be limited to a small fixed number of target species sets in order to ensure that accuracy and efficiency are maintained even for unanticipated combustion conditions.
\item The method will be generally applicable, and easy to extend to new problem types and reaction mechanisms.
\item The method will reduce the required user expertise necessary to determine appropriate DRGEP target species.
\end{enumerate}

\section{The Relative Importance Index Method}
\label{sec:RII}

\subsection{Sample Ignition Studies}
In order to develop this new method of dynamic target species selection, two constant-volume autoignition studies will be examined using version 2 of the LLNL detailed n-heptane mechanism~\cite{nhept_1,nhept_2,nhept_website}.
The sample cases will be taken at initial conditions of \SI{700}{\kelvin} and \SI{1000}{\kelvin}, $\phi = 1$, and \SI{2}{\atm} in order to explore the effects of different dominant combustion processes (low-temperature and high-temperature combustion, respectively).
The low-temperature case additionally exhibits a two-stage ignition process.
This provides an opportunity to investigate the effect of initial temperature and changing chemical pathways on the developed method.

\subsection{Graph Structure of the Directed Relation Graph with Error Propagation}
\label{sec:columnsum}

\begin{figure}
\centering
\begin{subcaptionbox}{An example diagram of a species and a neighbor in the DRGEP graph structure.  The dark and light grey arrows correspond to the lines in \subref{fig:ex_matrix}, i.e. to the rows and columns of the adjacency matrix respectively.  Species $B$ corresponds to a representative neighbor of $A$, and is marked on the adjacency matrix.\label{fig:ex_graph}}[0.5\linewidth]{
	\centering
	\includegraphics[width = 0.5\linewidth, keepaspectratio = true]{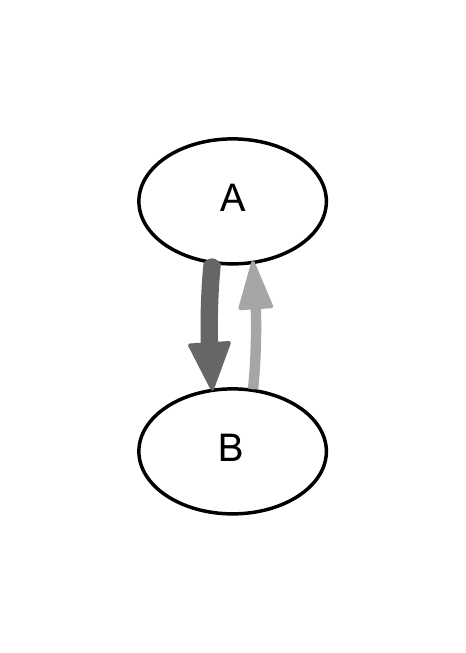}
}
\end{subcaptionbox} \\
\begin{subcaptionbox}{The adjacency matrix\label{fig:ex_matrix}}[0.5\linewidth]{
	\centering
	\includegraphics[width = 0.5\linewidth, keepaspectratio = true]{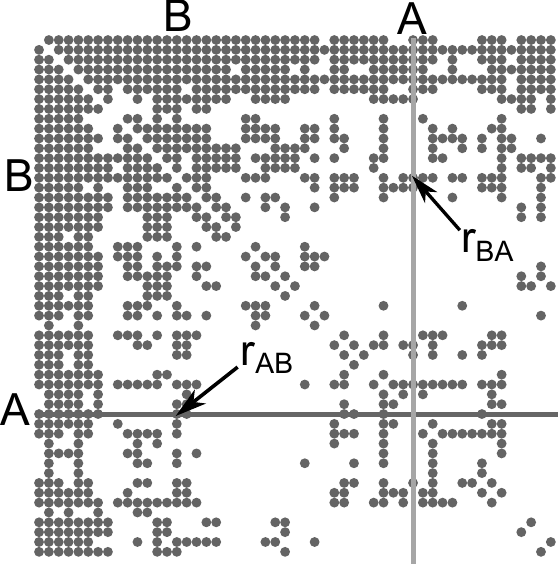}
}
\end{subcaptionbox}
\caption{An example of \subref{fig:ex_graph} the DRGEP graph and \subref{fig:ex_matrix} adjacency matrix structure used in this work created using the GRI 3.0 mechanism~\cite{gri30}.}
\label{fig:matrix}
\end{figure}

The DRGEP adjacency matrix in this work is formatted in row-major format: the value at row $i$, column $j$ is the DRGEP direct interaction coefficient (DIC) representing the dependence of species $i$ on species $j$, i.e.,\ \DIC{ij}.
Figure~\ref{fig:matrix} shows an example adjacency matrix generated for the GRI 3.0 mechanism~\cite{gri30}.
A species $A$ is considered a neighbor of species $B$ if they both participate in at least one reaction together.

The goal of \DIC{AB} is to quantify the dependence of species $A$ on $B$; therefore, if \DIC{AB} is large, $B$ is considered important to the production or consumption of $A$.
We consider the sum of the DICs of the neighbors of species $A$ in column A (light grey line in Fig.~\ref{fig:matrix}):

\begin{equation}
\text{Column Sum(A)} = \sum_{B_i \in \set{\text{neighbors(A)}}} \DICeq{B_iA} \;.
\label{eq:colsum}
\end{equation}

\begin{figure}
\begin{subcaptionbox}{\SI{700}{\kelvin} ignition case \label{fig:riicol_a}}{
	\centering
	\includegraphics[keepaspectratio = true]{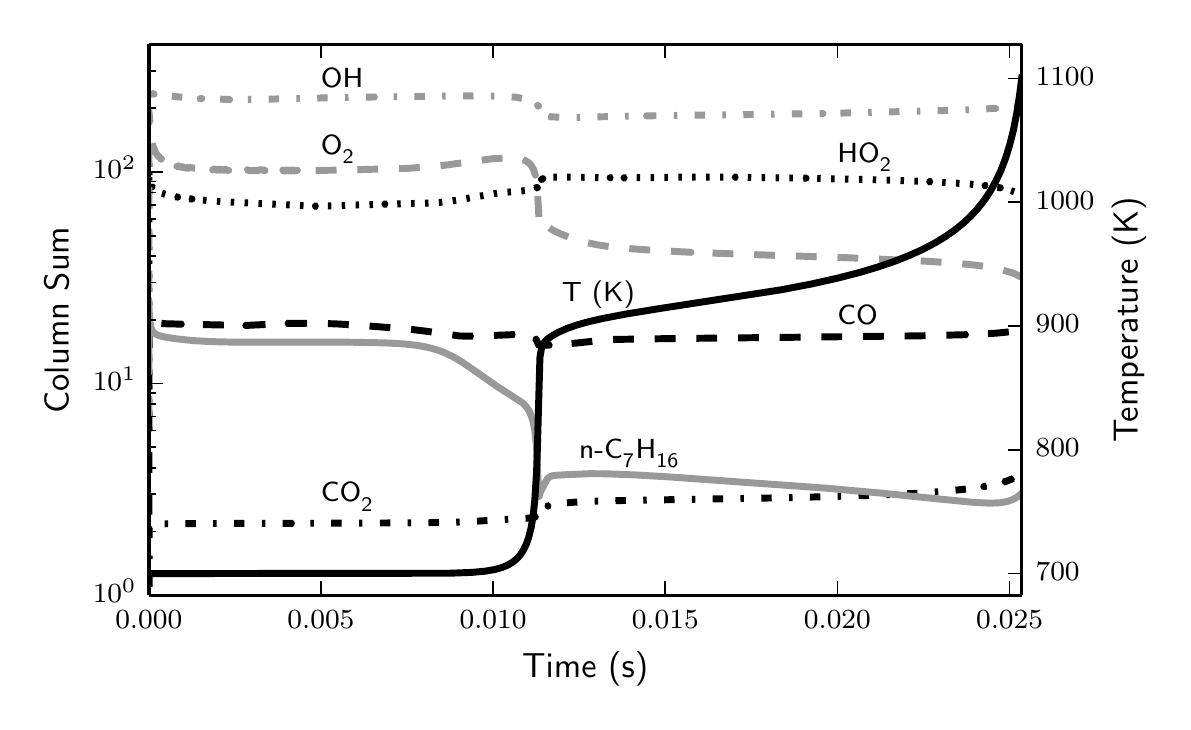}
}
\end{subcaptionbox}
\begin{subcaptionbox}{\SI{1000}{\kelvin} ignition case \label{fig:riicol_b}}{
	\centering
	\includegraphics[keepaspectratio = true]{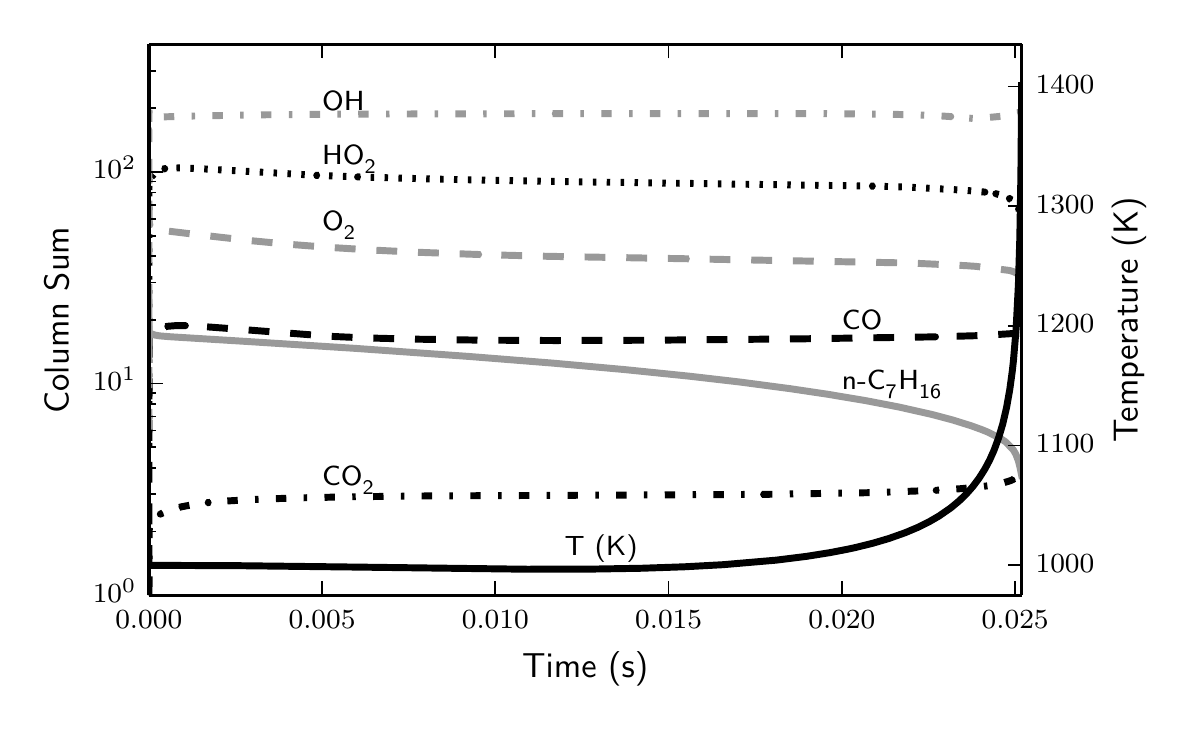}
}
\end{subcaptionbox}
\caption{The Column Sum of commonly used target species for the sample n-heptane constant-volume ignition cases.}
\label{fig:riicol}
\end{figure}

A species with a large column sum will be important to the production or consumption of many other species, and therefore will be very active in the mechanism.
However, since the column sum of a species depends on its number of neighbors, the column sum cannot be used directly to make a fair comparison between different species.
Yet, a sense of its utility in tracking the activity of a species can be observed by comparing column sums of the same species in different ignition cases, as depicted in Fig.~\ref{fig:riicol}.

In the \SI{700}{\kelvin} ignition case (Fig.~\ref{fig:riicol_a}), much of the n-heptane is consumed during first-stage ignition; as a result, the column sum of n-heptane drops off significantly at this point.
In addition, several large fuel fragments (e.g.,\ \ce{i{\hyphen}C3H6CO}, \ce{C4H7CO{1\hyphen4}}, \ce{n{\hyphen}C5H11CO}; see \cite{nhept_website} for species dictionary) decompose almost entirely in \ce{CO}-forming reactions; these pathways shut off near first-stage ignition and are responsible for the corresponding drop in the column sum of \ce{CO} (note: the magnitude of this drop is masked somewhat by use of a log scale).
In the low-temperature region before first-stage ignition, the column sums of \ce{O2} and \ce{OH} are particularly large as a result of the enhanced \ce{R-OH} and \ce{R-O2} chemistry.
The column sum of \ce{HO2} is lower in this region compared to its column sum after first-stage ignition, indicating it is less important to the mechanism at this lower temperature state.
The column sum of \ce{CO2} is small in the low-temperature region, as it is only formed through a handful of reactions.
After first-stage ignition, the column sums of all species except n-heptane reach levels similar to that of the \SI{1000}{\kelvin} case, indicative of the temperature effect on the changing strength of chemical pathways.

The absence of a multistage ignition event in the \SI{1000}{\kelvin} ignition case (Fig.~\ref{fig:riicol_b}) manifests in far more gradual changes in the column sums.
The column sum of n-heptane steadily declines as it is consumed.
As in the \SI{700}{\kelvin} ignition case, the column sum of \ce{CO} is initially bolstered due to the decomposition of large fuel fragments (e.g.,\ \ce{a{\hyphen}C3H5CO}, \ce{i{\hyphen}C3H7CO}, \ce{C4H7CO{1\hyphen4}}) forming \ce{CO}; as these reactions slow down, the column sum of \ce{CO} declines, becoming stable until climbing slowly leading up to ignition as \ce{CO} becomes important for \ce{CO2} creation.
The column sum of \ce{OH} stays roughly constant throughout ignition, while the column sum of \ce{HO2} declines approaching ignition as one of its primary consumption pathways, via reactions with n-heptane, slows down.
Finally, the column sum of \ce{CO2} is fairly constant until increasing leading up to ignition as \ce{CO2} formation increases. 

\subsection{Target Species Roles}
\label{sec:roles}

\begin{figure}
\begin{subcaptionbox}{\ce{n{\hyphen}C7H16}}[0.5\linewidth]{
	\centering
	\includegraphics[width=0.5\linewidth, keepaspectratio = true]{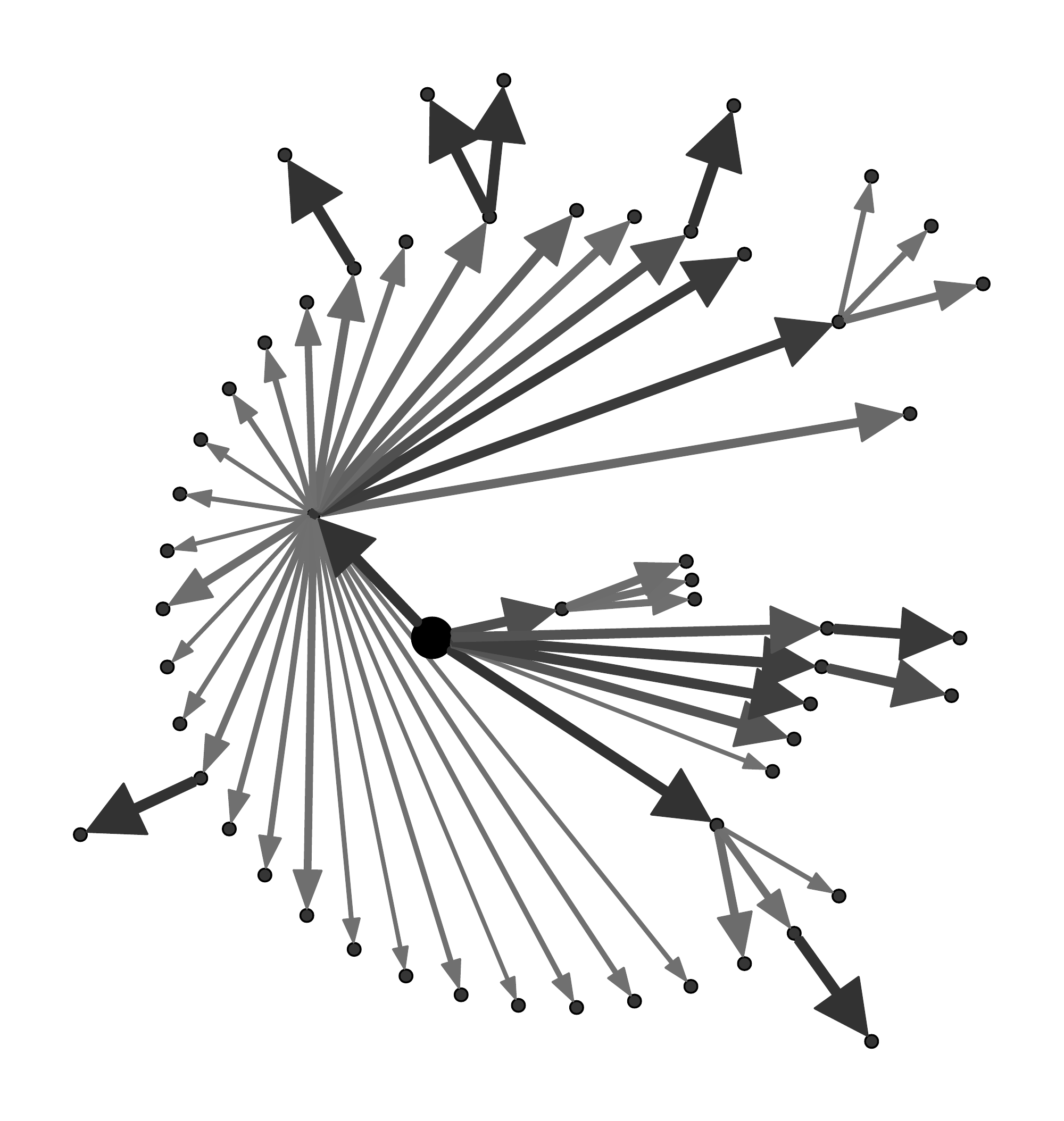}
}
\end{subcaptionbox}
\begin{subcaptionbox}{\ce{OH}}[0.5\linewidth]{
	\centering
	\includegraphics[width=0.5\linewidth, keepaspectratio = true]{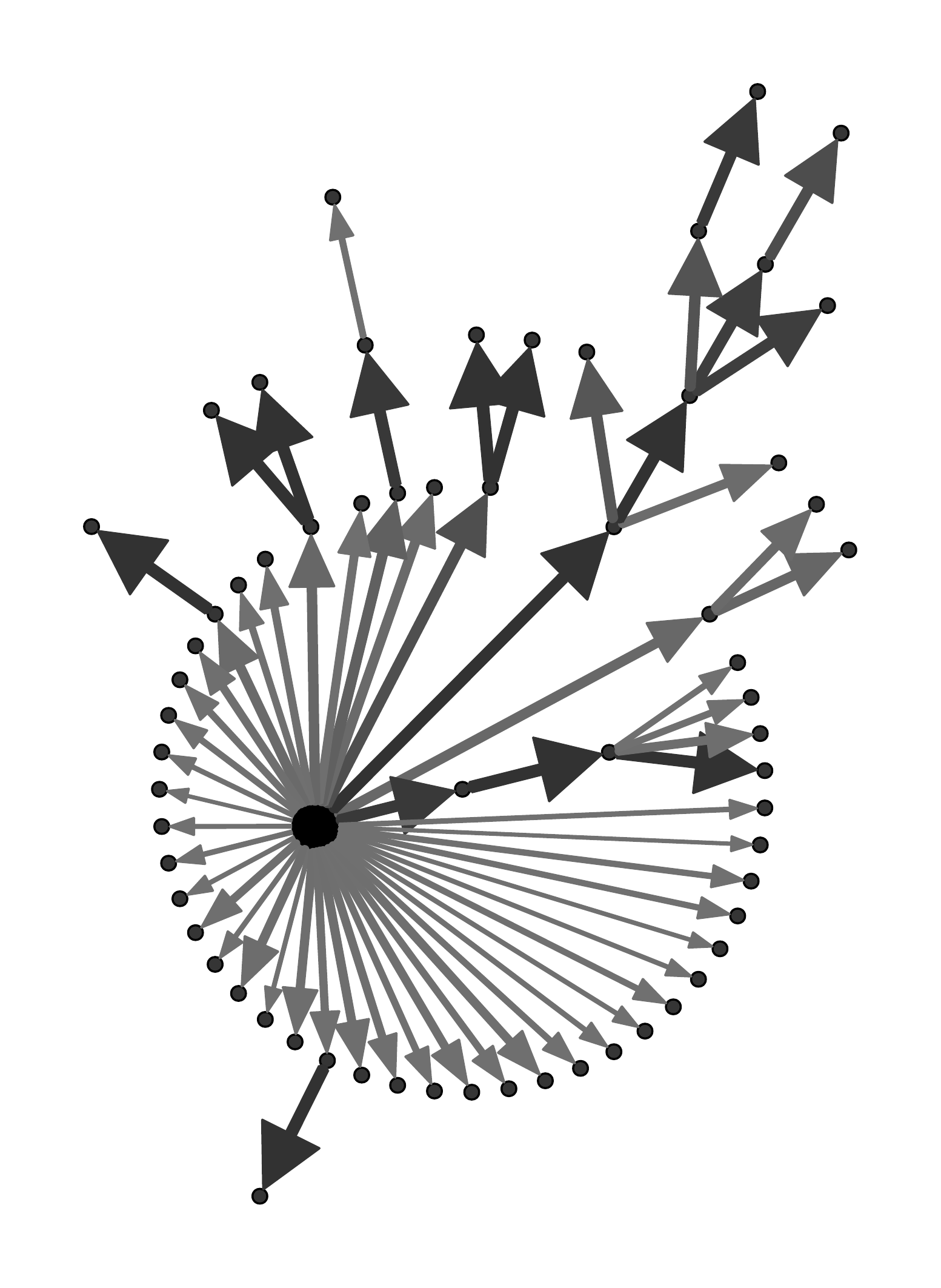}
}
\end{subcaptionbox}
\caption{A sample DRGEP reduction with $\Edrgep = $ \num{e-5} for two choices of target species.  The large black circles indicate (a) \ce{n{\hyphen}C7H16} and (b) \ce{OH}.  Arrow width and color indicate the magnitude of the direct interaction coefficient between species.}
\label{fig:local reduction}
\end{figure}

Common choices of static target species tend to fall into one of two categories: important radicals such as \ce{OH} and \ce{HO2}, and important reactants\slash products such as the fuel, \ce{O2}, \ce{CO2}, and \ce{CO}.
Figure~\ref{fig:local reduction} presents a sample DRGEP reduction of the n-heptane mechanism using either \ce{OH} or n-heptane as the sole target species, with $\Edrgep =$ \num{e-5}.

By examining the neighbors of n-heptane on the reduction graph, it becomes clear that n-heptane only directly participates with a handful of species, but tends to be strongly dependent on each.
Most species included in the resultant skeletal mechanism are kept due to their importance to the first ring of species strongly linked to n-heptane.
A species behaving in this manner will be called a ``locally important'' target species, due to its strong but localized link to a few species.
In this mechanism, \ce{CO}, \ce{CO2}, and n-heptane all behave as locally important target species.

The \ce{OH} reduction graph exhibits an almost entirely different pattern: most of the species included are one step away from \ce{OH} on the graph with fairly weak direct interaction coefficient links.
As a result, relatively few species not directly neighboring \ce{OH} are included by its choice as a target species.
Further, the included species that are more than one step away on the graph from \ce{OH} tend to be strongly linked to species that directly neighbor \ce{OH}.
This type of behavior will be termed a ``globally important'' target species, due to its direct but weak link to many species.
It has been found that most radicals (e.g.,\ \ce{OH}, \ce{HO2}) behave as globally important target species; in addition, non-radical species involved in many reactions (e.g.,\ \ce{O2}) tend to behave as globally important targets as well. 

Whether a species behaves as a locally or globally important target is largely controlled by its number of neighbors---i.e., the number of species it directly interacts with through reactions---in the mechanism.
Table~\ref{tab:neighbors} lists the number of neighbors and corresponding reactions for each of the discussed target species.
In general, global target species will tend to have many neighbors in a reaction mechanism.
By examining Eq.~\eqref{eq:colsum}, it can be inferred that globally important target species will have larger column sums than locally important target species in general; as such, the column sum alone is insufficient to compare potential target species.

\subsection{Column Sum Normalization Using Row Average}
\label{sec:riirow}

The magnitude of the column sum of a species will roughly depend on its number of neighbors in a mechanism.  
Therefore, a normalizing factor that tends to be smaller for species with many active neighbors will be necessary.
In the n-heptane mechanism used, the \ce{OH} radical participates in 1051 reactions with 519 different species, while n-heptane is involved in 146 reactions with 36 species as shown in Table~\ref{tab:neighbors}.
\ce{OH} reacts with so many different species that most neighboring species will only be involved in a relatively small fraction of \ce{OH}'s total production and consumption.
However, n-heptane reacts with relatively fewer species, meaning that most of its neighboring species may control an appreciable fraction of its total consumption and production.
From Eq.~\eqref{eq:DIC}, it can be inferred that the average DRGEP coefficient from \ce{OH} to its neighbors:
\begin{equation}
\overline{\DICeq{\ce{OH}}} = \frac{\sum_{B_i\in\set{\text{neighbors(\ce{OH})}}} \DICeq{\ce{OH},B_i}}{\sum_{B_i\in\set{\text{neighbors(\ce{OH})}}} \delta_{\ce{OH},B_i}} \;,
\end{equation}
where
\begin{equation}
\delta_{\ce{OH},B_i} = \begin{cases}
1 & \text{if } \DICeq{\ce{OH},B_i} > 0 \text{ and}\\
0 & \text{if } \DICeq{\ce{OH},B_i} = 0 \;,
\end{cases}
\end{equation}
will be much smaller than that of n-heptane, i.e.,:
\begin{equation}
\overline{\DICeq{\ce{OH}}} \ll \overline{\DICeq{\ce{n{\hyphen}C7H16}}} \;.
\label{eq:rowavg-pred}
\end{equation}
Consequently, the Row Average of a species $A$ is defined as:
\begin{equation}
\text{Row Average(A)} = \frac{\sum_{B_i\in\set{\text{neighbors(A)}}} \DICeq{AB_i}}{\sum_{B_i\in\set{\text{neighbors(A)}}} \delta_{AB_i}} \;.
\end{equation}

\begin{figure}
	\begin{subcaptionbox}{\SI{700}{\kelvin} ignition case}{
		\centering
		\includegraphics[keepaspectratio = true]{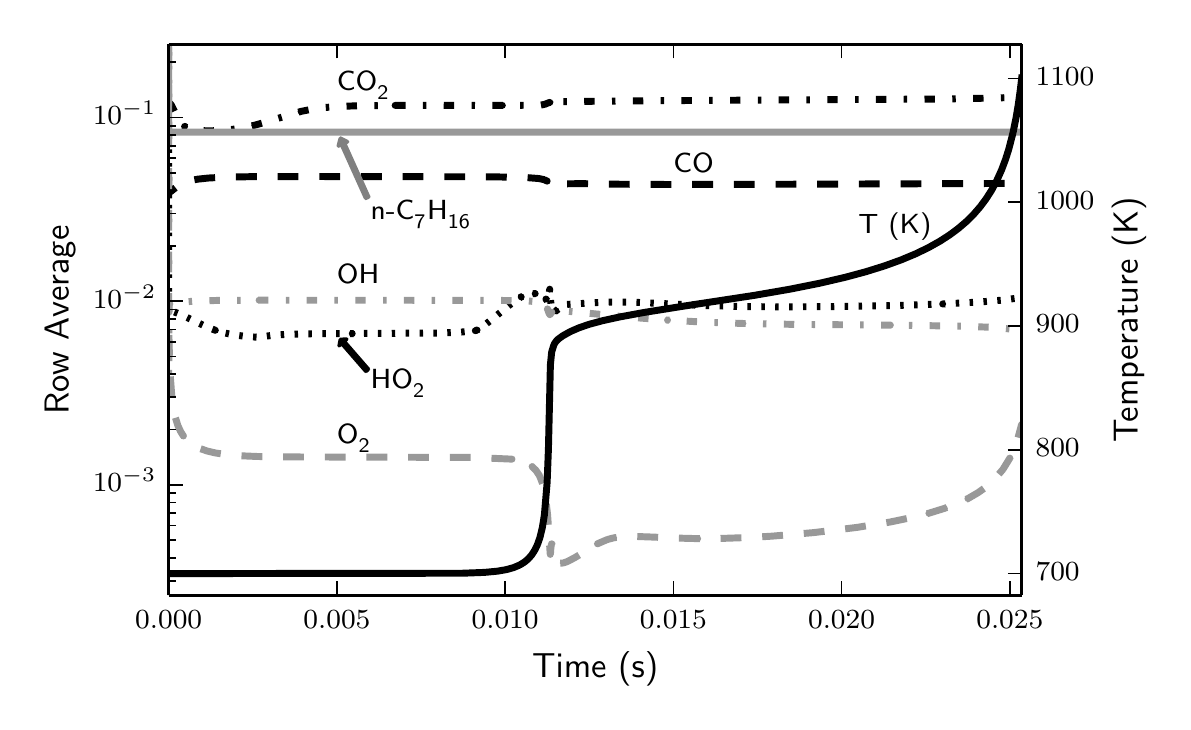}
	}
	\end{subcaptionbox}
	\begin{subcaptionbox}{\SI{1000}{\kelvin} ignition case}{
		\centering
		\includegraphics[keepaspectratio = true]{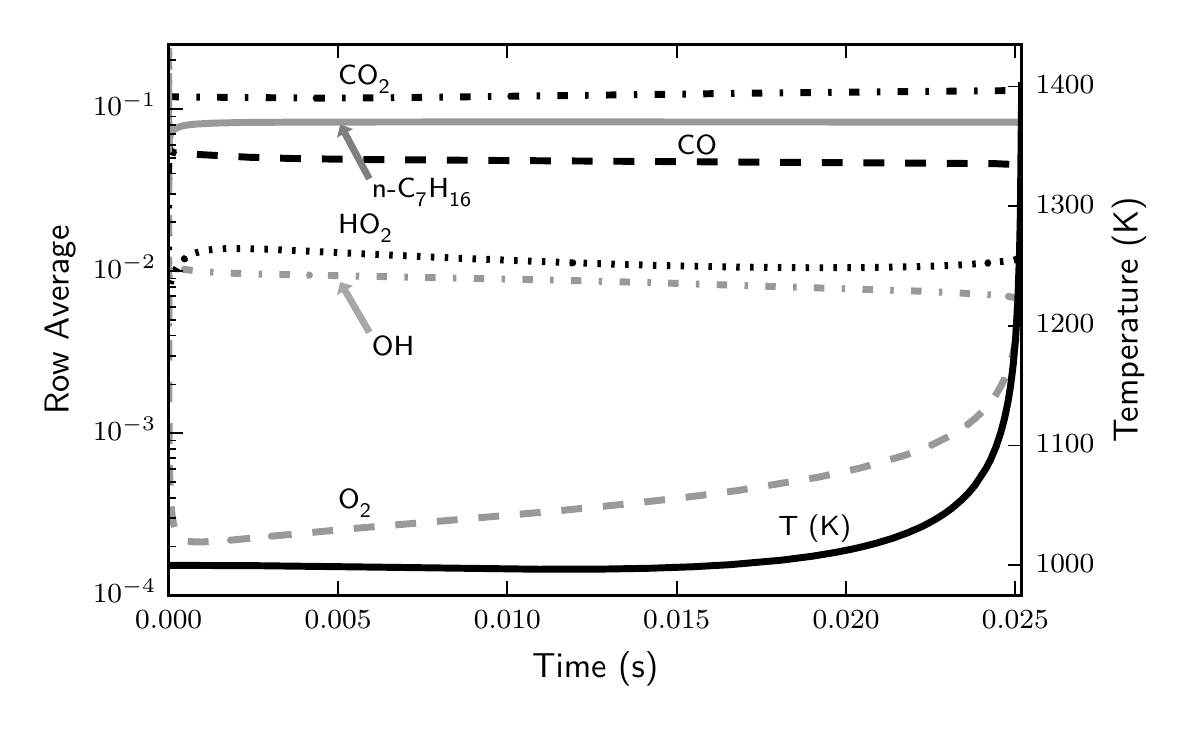}
	}
	\end{subcaptionbox}
	\caption{The Row Average of commonly used target species for the sample n-heptane constant-volume ignition cases.}
	\label{fig:riirow}
\end{figure}

Figure~\ref{fig:riirow} shows the row averages for common target species, where it is seen that the relationship predicted by Eq.~\eqref{eq:rowavg-pred} holds; in both ignition cases, the row average of \ce{OH} is almost an order of magnitude smaller than that of n-heptane.
Further, the species expected to be locally important targets (e.g.,\ \ce{CO}, \ce{n{\hyphen}C7H16}, \ce{CO2}), as listed in Table~\ref{tab:neighbors}, have row averages around an order of magnitude larger than the globally important target species (e.g.,\ \ce{O2}, \ce{OH}, \ce{HO2}) for both ignition cases.

\subsection{Definition of the Relative Importance Index}
\label{sec:rii_def}

Using the row average as a normalizing factor, the relative importance index (RII) is defined as:
\begin{equation}
\text{RII(A)} = \text{Row Average(A)} \times \text{Column Sum(A)}
\end{equation}

\begin{figure}
	\begin{subcaptionbox}{\SI{700}{\kelvin} ignition case\label{fig:rii_a}}{
		\centering
		\includegraphics[keepaspectratio = true]{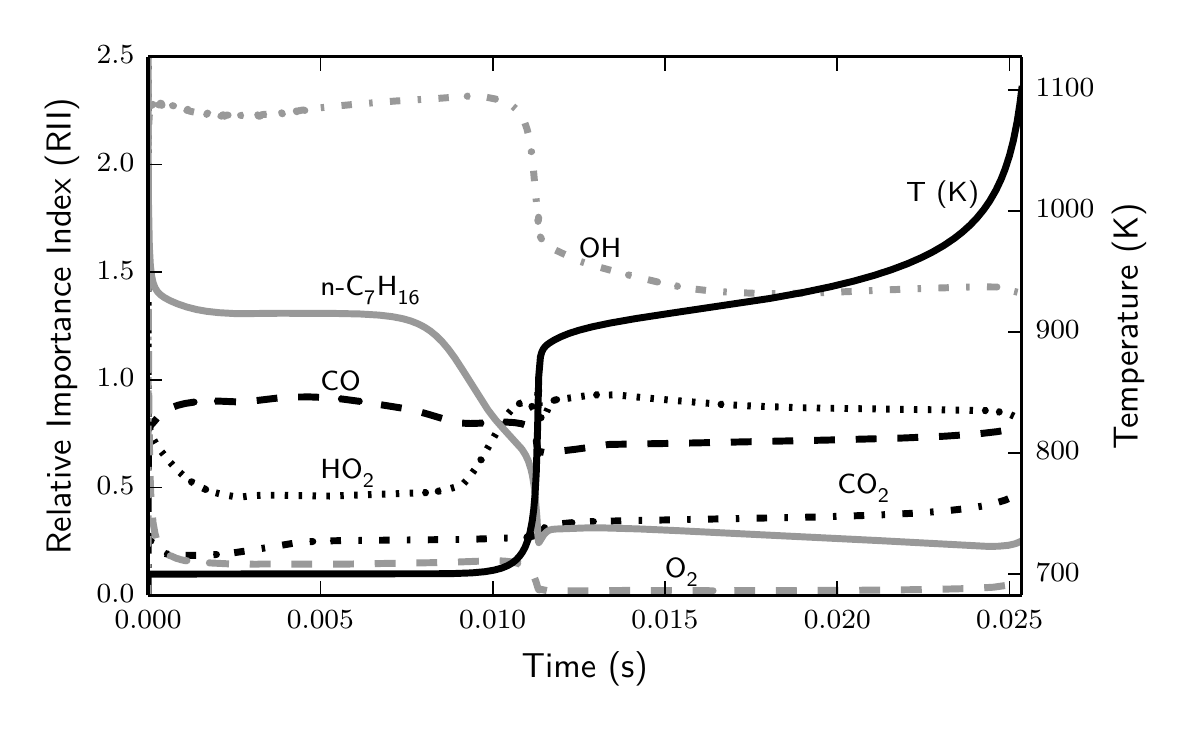}
	}
	\end{subcaptionbox} \\
	\begin{subcaptionbox}{\SI{1000}{\kelvin} ignition case\label{fig:rii_b}}{
		\centering
		\includegraphics[keepaspectratio = true]{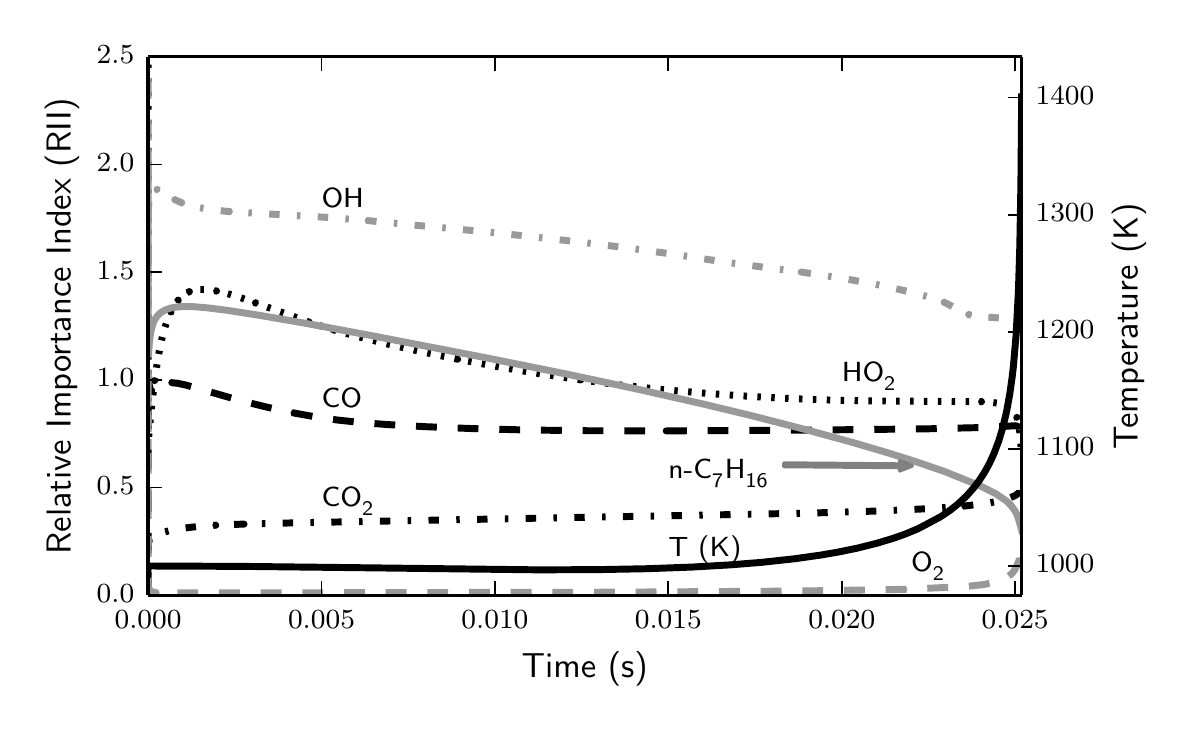}
	}
	\end{subcaptionbox}
	\caption{The RII of commonly used target species for the sample n-heptane constant-volume ignition cases.}
	\label{fig:rii}
\end{figure}

Figure~\ref{fig:rii} examines the RII values of commonly used target species for the sample ignition cases.
In both cases, the RII of \ce{OH} remains one of the largest throughout the simulation. 
In the \SI{700}{\kelvin} ignition case (Fig.~\ref{fig:rii_a}), the RII of n-heptane declines as it begins to be consumed during first-stage ignition, while the RII of \ce{HO2} rises as new pathways open leading up to the first ignition event.
In the low-temperature chemistry region, the RII of \ce{OH} and \ce{O2} are significantly higher than in the second-stage induction process, again due to the enhanced \ce{R-OH} and \ce{R-O2} activities.
The RII of \ce{CO} drops during first-stage ignition due to the closing of creation pathways, e.g., via the destruction of \ce{CH3CO} and \ce{C2H5CO}, as well as the destruction of large fuel fragments as discussed in Section~\ref{sec:columnsum}, and increases slowly leading up to the final ignition event as \ce{CO} becomes important for \ce{CO2} formation.
Finally, the RII of \ce{CO2} remains relatively low throughout, increasing slightly during first-stage ignition and continuing to increase leading up to the final ignition event. 

In the \SI{1000}{\kelvin} case (Fig.~\ref{fig:rii_b}), the RII values of \ce{HO2} and n-heptane match each other closely for the beginning of the induction period, as \ce{HO2} consumption is strongly coupled to n-heptane reactions; closer to the point of ignition, this phenomenon stops as \ce{HO2} begins to react with \ce{C2H4} and \ce{CO} instead.
Near the beginning of the simulation, \ce{OH} consumption and production are strongly tied to a few reactions (e.g.,\ formation of heptyl radicals via reactions with n-heptane), but as the system approaches the point of ignition, \ce{OH} consumption and production becomes weakly tied to reactions with many species; as a result, the RII of \ce{OH} drops slowly throughout the simulation.
Similar to its column sum, the RII of \ce{CO} declines as its creation via the decomposition of large fuel fragments slows down, and later increases slowly leading up to ignition.
Finally, \ce{CO2} has a small RII for most of the simulation, but slowly increases leading up to ignition, similar to the \SI{700}{\kelvin} case.

As demonstrated above, the RII of a species balances how strongly a species depends on the majority of its neighbors with how strongly its neighbors depend on it.
As such, the use of the RII method allows both globally important and locally important target species to be selected.

\subsection{Target Selection Process}
\label{sec:target selection}

The RII of each species can be calculated from the DRGEP coefficients computed during the reduction process.
Moreover, it is trivial to add the calculation of active neighbors, column sums, and row sums (later combined with the active neighbors to form the row average) to the pre-existing DRGEP calculation loop, meaning that this method can be executed with minimal overhead.
The only additional work required compared to the use of a static target set is a single loop that visits all of the species in the mechanism in order to put this information together to calculate the RII values.
In this loop, each species mass fraction is tested; if it is greater than a minimum threshold, the species is inserted into a priority queue, with priority equal to its RII value.
When the queue size becomes larger than the number of RII targets to be selected, the species with the minimum RII value is popped off the queue.
Depending on the priority queue implementation, the cost of each insertion/removal pair from the priority queue scales linearly or logarithmically with the maximum size of the queue (the number of RII target species to be selected).
However, since the number of RII target species is a small fixed number, the overall cost can be considered constant time in asymptotic analysis, i.e.,\ $\mathcal{O}(1)$.
Therefore, the total additional overhead of this method compared to using a static target species set scales linearly with the number of species in the mechanism: $\mathcal{O}(N_S)$.
As the asymptotic cost of the DRGEP method scales linearly with the number of reactions in the mechanism, $\mathcal{O}(N_R)$, and the number of reactions tends to scale as $N_R \approx 5 \times N_S$ \cite{lu_towardrealistic}, the overhead involved in calculating the RII values has a minimal impact on the overall speed of the mechanism reduction process.

Consider the hypothetical case where a species $A$ participates in only two reactions, one that produces $A$ and another that consumes $A$.
Species $A$ will naturally depend heavily on the few species it reacts with, and have a relatively large row average (compared to those shown in Fig.~\ref{fig:riirow}) as a result.
If the neighboring species of $A$ also participate in relatively few reactions, then the DRGEP coefficients $\DICeq{B_iA} \text{ for } B_i \in \set{\text{neighbors(A)}}$ will tend to be closer to \num{1.0}, as each neighbor $B_i$ is likely to depend strongly on $A$.
The resulting column sum of $A$ will then be reasonably similar in magnitude to the number of neighbors of $A$.
As discussed in Section~\ref{sec:rii_def}, the RII responds to how active a species is in various parts of a simulation; however, for such a species $A$ the RII may be large regardless.
In this case, it is difficult to determine the importance of $A$ from the RII value alone, so an additional criteria will be needed as a supplement.
An example detailing this case is presented in the appendix.

If species $A$ is only weakly reacting it will typically be present in small quantities, whereas when it is strongly reacting it will tend to be present in much larger quantities.
Therefore, applying a mass fraction threshold in combination with the RII value can help to properly identify the importance of species.
For example, by using a relatively large mass fraction threshold (e.g., \num{e-8}), such a species $A$ will not be considered unless it is strongly reacting.
Although a more rigorous filtering method may be more effective in eliminating these potentially biased RII values, this simple combined method proves effective in our studies.

\begin{figure}
	\begin{subcaptionbox}{First stage ignition \label{subfig:700K-first}}[\linewidth]{
		\centering
		\includegraphics[width = 0.75\linewidth, keepaspectratio = true]{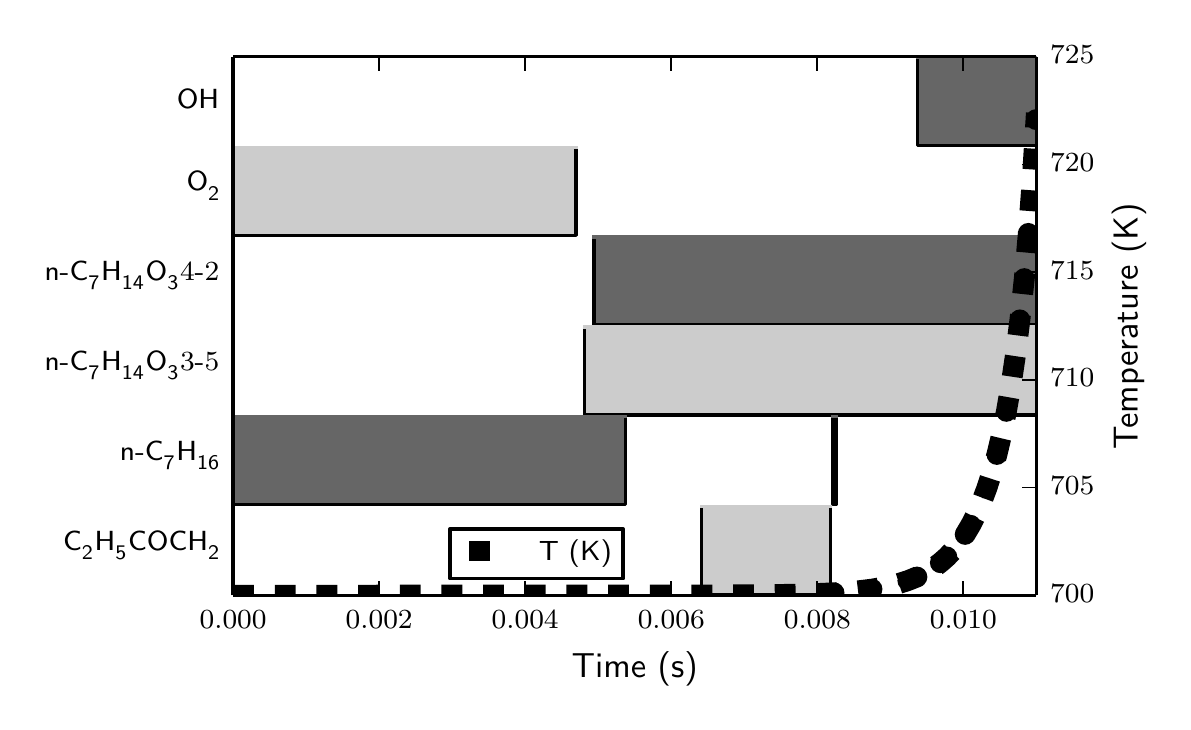}
	}
	\end{subcaptionbox}\\
	\begin{subcaptionbox}{Second stage ignition \label{subfig:700K-second}}[\linewidth]{
		\centering
		\includegraphics[width = 0.75\linewidth, keepaspectratio = true]{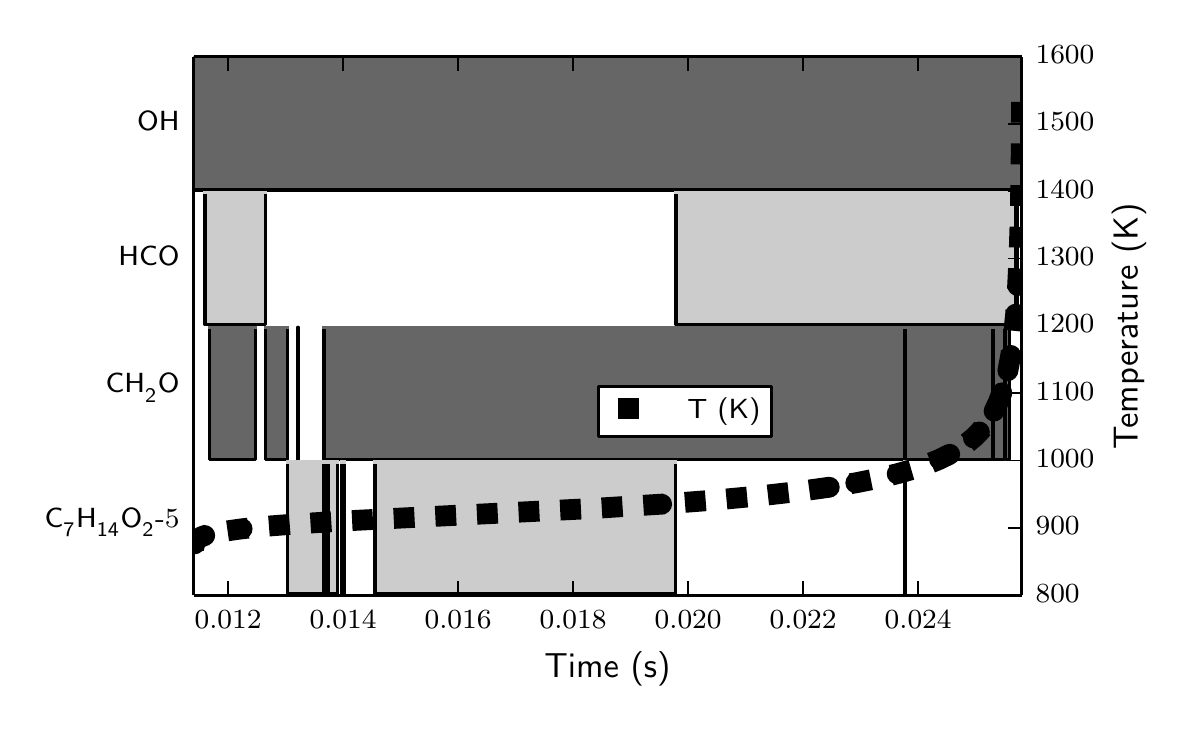}
	} 
	\end{subcaptionbox}\\
	\begin{subcaptionbox}{Post ignition \label{subfig:700K-post}}[\linewidth]{
		\centering
		\includegraphics[width = 0.75\linewidth, keepaspectratio = true]{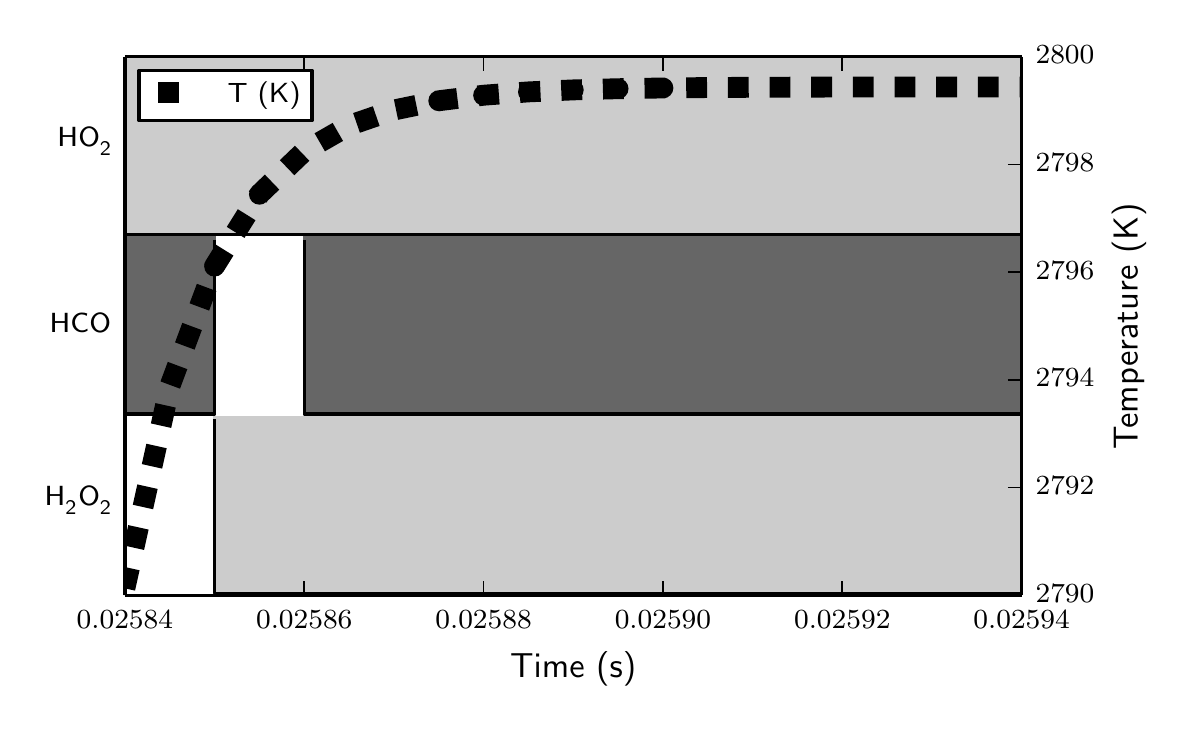}
	}
	\end{subcaptionbox}
	\caption{Target species selection for a three RII target species criterion with mass fraction cutoff of \num{e-8} for the \SI{700}{\kelvin} ignition case exhibiting two-stage ignition response.}
	\label{fig:700K target selection}
\end{figure}

For the \SI{700}{\kelvin} ignition case, Fig.~\ref{fig:700K target selection} examines the target species selected by the RII method with three RII target species and a mass fraction cutoff of \num{e-8}.
During first-stage ignition (Fig.~\ref{subfig:700K-first}), n-heptane and \ce{O2} are initially selected as target species simply because they are the reactants.
Around \SI{0.005}{\second}, two seven carbon ketones (\ce{n{\hyphen}C7H14O3{3\hyphen5}} and \ce{n{\hyphen}C7H14O3{4\hyphen2}}) are selected as target species by the RII method; these ketones play a large role in \ce{OH} creation in this region, via both their creation and destruction, and lay along major pathways in the low-temperature breakdown of n-heptane.
At approximately \SI{0.007}{\second}, \ce{C2H5COCH2}---a major product of the decomposition of these ketones---begins to be selected as a target; \ce{C2H5COCH2} additionally lies along an important path for early \ce{CO2} production, forming a bulk of the \ce{CH2CO} and \ce{C2H5} in the system.
Once the concentration of \ce{OH} builds up, it is continually selected until the final ignition event.
Radicals such as \ce{OH} and other species that participate in many reactions in the mechanism are often important even in small concentrations.
Further, not selecting these species as targets can greatly increase error, e.g., as in the low-temperature ignition example shown in Fig.~\ref{fig:conv static err} where selecting \ce{OH} as a target species significantly improved accuracy.
For this reason, a relaxed mass fraction cutoff of \num{e-15} was used for any species that participated in more than \SI{10}{\percent} of the reactions in the mechanism or neighbored more than \SI{50}{\percent} of the species in the mechanism.
Applying this relaxed mass fraction cutoff results in the selection of \ce{OH} as a RII target species near the beginning of the simulation.

During second-stage ignition (Fig.~\ref{subfig:700K-second}), the \ce{HCO} created during the first stage begins to react, and it is selected as a target species until \ce{HCO} levels are depleted.
Around the same time (approximately \SI{0.012}{\second}), formaldehyde (\ce{CH2O})---an important \ce{HCO} precursor---begins to be selected as a target, and will continue to be selected for most of the second stage induction period.
The selection of \ce{C7H14O2{\hyphen}5} as a target species is an example of the biased RII values discussed previously (and detailed in the appendix).
Although involved in the breakdown of the leftover large fuel fragments from first-stage ignition, this chemistry is of relatively minor importance to the mechanism at this point and therefore \ce{C7H14O2{\hyphen}5} would likely not be considered as a target species given a more rigorous RII bias detection method.
After the final ignition event the chemistry is dominated by the \ce{H2-O2} and \ce{CO} oxidation reactions, and hence \ce{HO2}, \ce{HCO}, and \ce{H2O2} are chosen as target species.

\begin{figure}
	\begin{subcaptionbox}{Ignition \label{subfig:1000K-pre}}[\linewidth]{
		\centering
		\includegraphics[width = 0.75\linewidth, keepaspectratio = true]{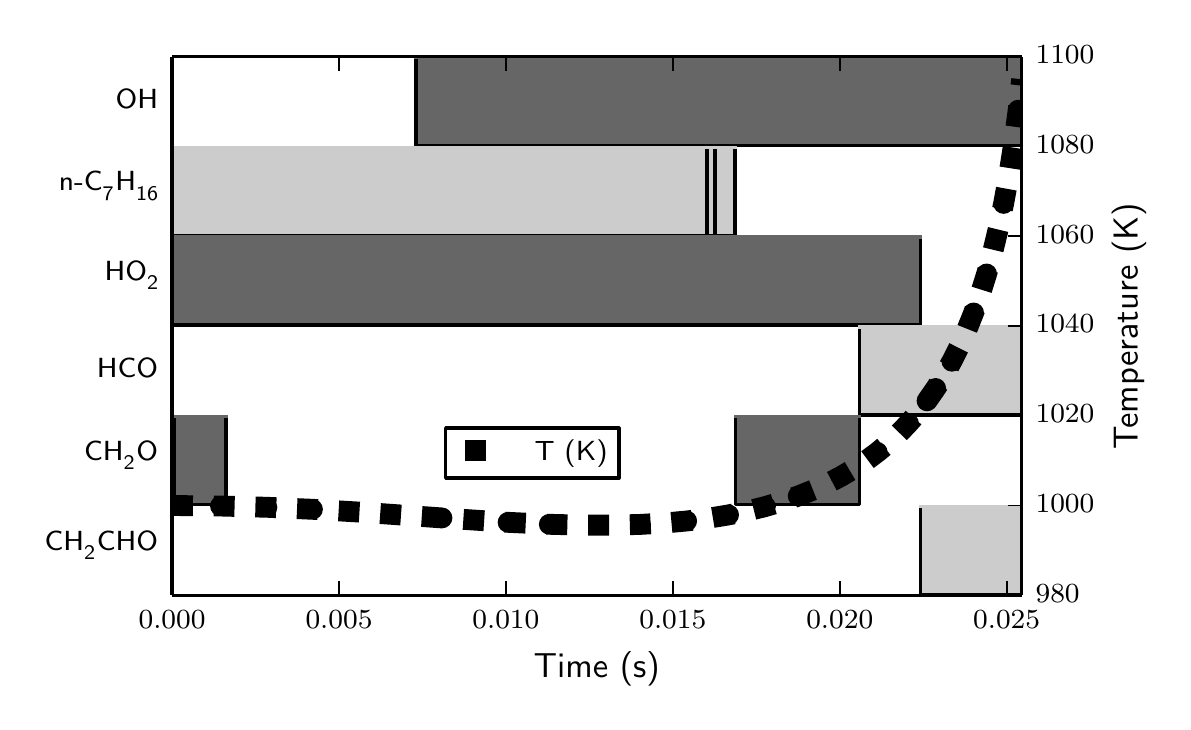}
	}
	\end{subcaptionbox}\\
	\begin{subcaptionbox}{Post ignition \label{subfig:1000K-post}}[\linewidth]{
		\centering
		\includegraphics[width = 0.75\linewidth, keepaspectratio = true]{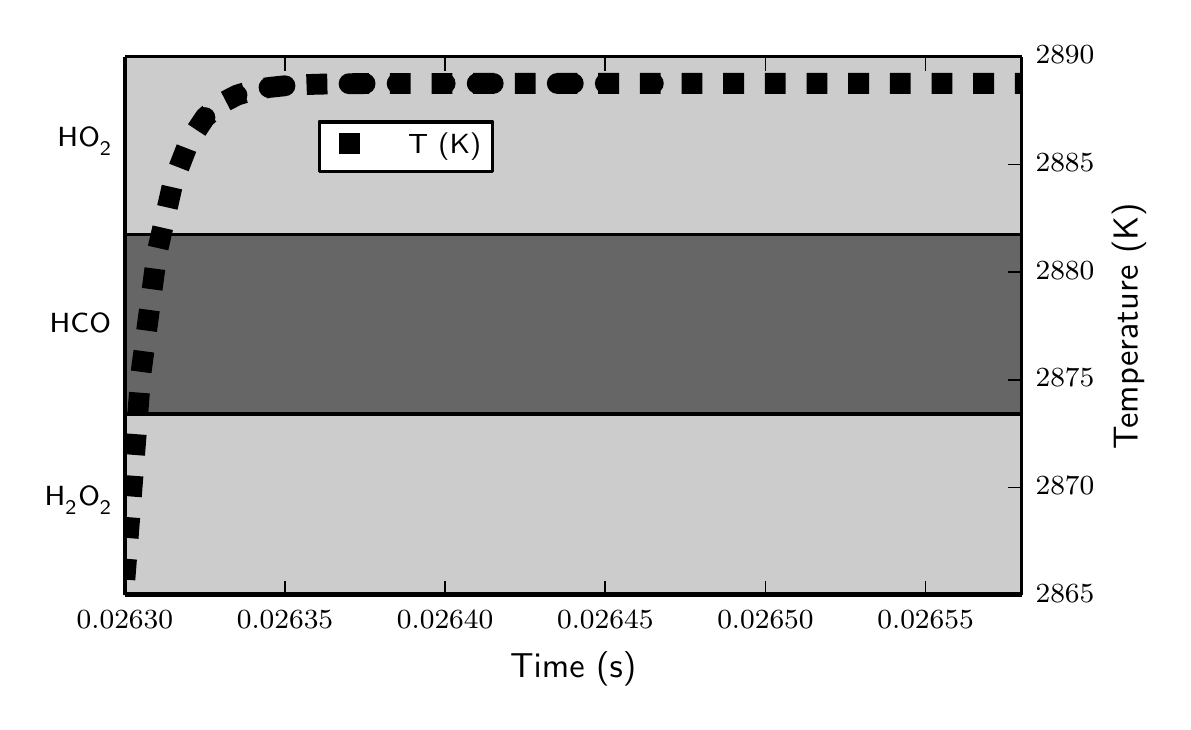}
	}
	\end{subcaptionbox}
	\caption{Target species selection for a three RII target species criterion with mass fraction cutoff of \num{e-8} for the \SI{1000}{\kelvin} ignition case exhibiting single-stage ignition response.}
	\label{fig:1000K target selection}
\end{figure}

In the \SI{1000}{\kelvin} ignition case, shown in Fig.~\ref{fig:1000K target selection}, n-heptane is again initially selected as a target species.
At this higher temperature, destruction of n-heptane is initially dominated by reactions with \ce{HO2}.
In addition, \ce{HO2} is the primary producer of \ce{H2O2} and an important secondary creation source for \ce{OH} through reactions with \ce{CH3} and \ce{C2H5}, resulting in the selection of \ce{HO2} as a target species for much of the induction period.
Initially, a large portion of \ce{CO} and \ce{OH} production occurs via \ce{CH2O} formation reactions, resulting in its brief selection as a target species; closer to ignition, \ce{CH2O} is selected again as it becomes involved in the two primary \ce{HCO} production pathways.
Shortly thereafter, \ce{CH2O} is replaced as a target species by \ce{HCO} as it becomes the primary producer of \ce{CO} and \ce{HO2}, while becoming the primary consumer of \ce{OH}.
Near ignition, \ce{CH2CHO} begins to be selected as a target species; as the primary product of the decomposition of several large ketones (e.g.,\ \ce{n{\hyphen}C7H14O3{1\hyphen3}}, \ce{n{\hyphen}C5H10O3{1\hyphen3}}), \ce{CH2CHO} reacts almost exclusively with oxygen, to form \ce{CH2O}, \ce{OH}, and \ce{CO}.
In addition to being an important secondary path for the creation of \ce{CH2O}, it provides an important path to the \ce{CO2} creation reactions.
After ignition, the same targets were selected as in the \SI{700}{\kelvin} case post ignition: \ce{HO2}, \ce{HCO}, and \ce{H2O2}, covering \ce{H2-O2} and \ce{CO} oxidation reactions.

\subsection{Adaptability to Combustion Conditions}
\label{sec:adapt}
\begin{figure}
\centering
\begin{subfigure}[t]{\linewidth}
	\centering
	\includegraphics[width = \linewidth, keepaspectratio = true]{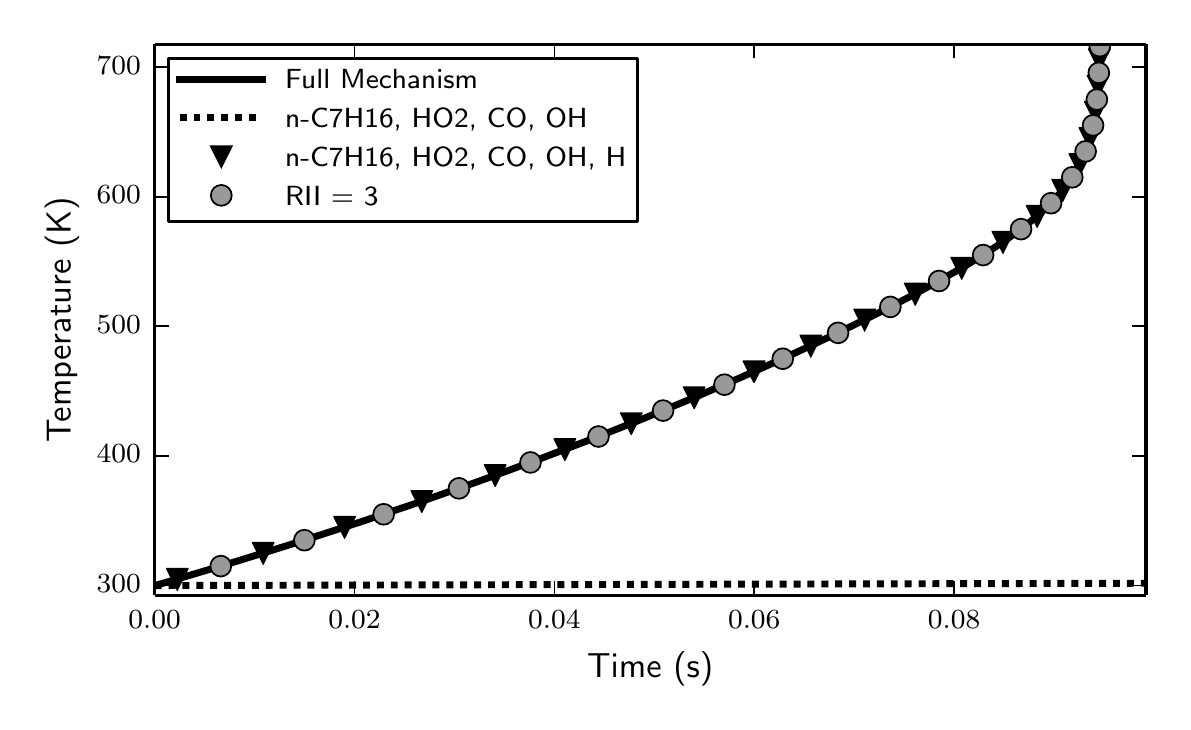}
\end{subfigure}
\caption{Temperature trace of two static target species sets compared to that of a three RII target species criteria for the constant-pressure reactor ignition of a stoichiometric n-heptane\slash air mixture by a pilot stream of the \ce{H} radical.  The mass flow rate of the pilot stream is 1:100 with that of the n-heptane\slash air mixture.}
\label{fig:diffusion-dynamic}
\end{figure}

In Section~\ref{sec:Applicability}, it was found that a static target species set not containing the \ce{H} radical could not capture the ignition of a stream of stoichiometric n-heptane/air flowing into a constant-pressure reactor by a pilot stream of the \ce{H} radical (Fig.~\ref{fig:diffusion-static}).  
In Fig.~\ref{fig:diffusion-dynamic}, the temperature traces of a three RII target species criterion are compared to those of the static target sets for the same problem.
As discussed in Section~\ref{sec:target selection}, highly connected species were identified as species participating in \SI{10}{\percent} or more of the total reactions, or neighboring \SI{50}{\percent} of total species.  
For this mechanism, the species identified by these criteria were \ce{H}, \ce{HO2}, \ce{O2}, and \ce{OH} and a relaxed mass fraction cutoff of \num{e-15} was used for each.
The RII target species criteria has no difficulty picking up the \ce{H} radical induced ignition, demonstrating an ability to adapt to different combustion conditions.
It is noted that when using a DRG-based DAC scheme in this example, the RII target species criteria can accurately predict ignition while a static target species set that does not include the H radical will rapidly accumulate large errors, similar to the DRGEP-based DAC scheme without considering the H radical. 
This further demonstrates that DRG-based DAC schemes will benefit from the use of an automated target selection method.

\section{Validation and Performance}
\label{Validation}

\subsection{n-Heptane Constant Volume Autoignition Simulations}
\label{sec:nhept-conv}

As seen in Section~\ref{sec:Applicability}, a target species set that is accurate for single-cell HCCI engine simulations can often be quite inaccurate in predicting constant-volume ignition delay.
Therefore, the accuracy of the RII method will first be assessed with constant-volume ignition delay predictions over a wide range of initial conditions in order to validate the method on a more stringent problem set.
The target sets were tested at \SI{5}{\atm} and \SI{20}{\atm}, equivalence ratios of $\phi = \set{0.5, 1, 2}$, and initial temperatures ranging from \SIrange{640}{1200}{\kelvin}.

Table~\ref{tab:nhept-conv-errorstats} compares the accuracy of target sets with three, four, and five RII targets to those of the static target sets \liangtargets\ and \fulltargets{}.
The mass fraction cutoffs were set to \num{e-30} for the static targets sets and \num{e-8} for the RII target species criteria.
As in Section~\ref{sec:adapt}, a relaxed mass fraction cutoff of \num{e-15} was used for the highly connected species \ce{H}, \ce{HO2}, \ce{O2}, and \ce{OH}.

It is seen from Table~\ref{tab:nhept-conv-errorstats} that the RII target criteria are capable of accurately predicting the ignition delays for the entire range of initial conditions.
Furthermore, the RII target species criteria are more accurate than the static target sets, having maximum errors of \SIrange{10.6}{5.45}{\percent} for \numrange{3}{5} RII targets compared to \SI{12.5}{\percent} for \fulltargets\ and \SI{22.5}{\percent} for \liangtargets{}.
It is noted that the maximum error does not decrease monotonically with increasing numbers of RII target species; the five RII target species criterion has a slightly greater maximum error than the four RII target species criterion.  However, the average error over all cases does decrease monotonically with larger RII target species criteria.

\subsection{n-Heptane HCCI Engine Simulations}

\begin{figure}
\begin{subcaptionbox}{$\Edrgep = $ \num{e-4} \label{subfig:nhept-hcci-lowerr}}
{
	\centering
	\includegraphics[width = 0.5\linewidth, keepaspectratio = true]{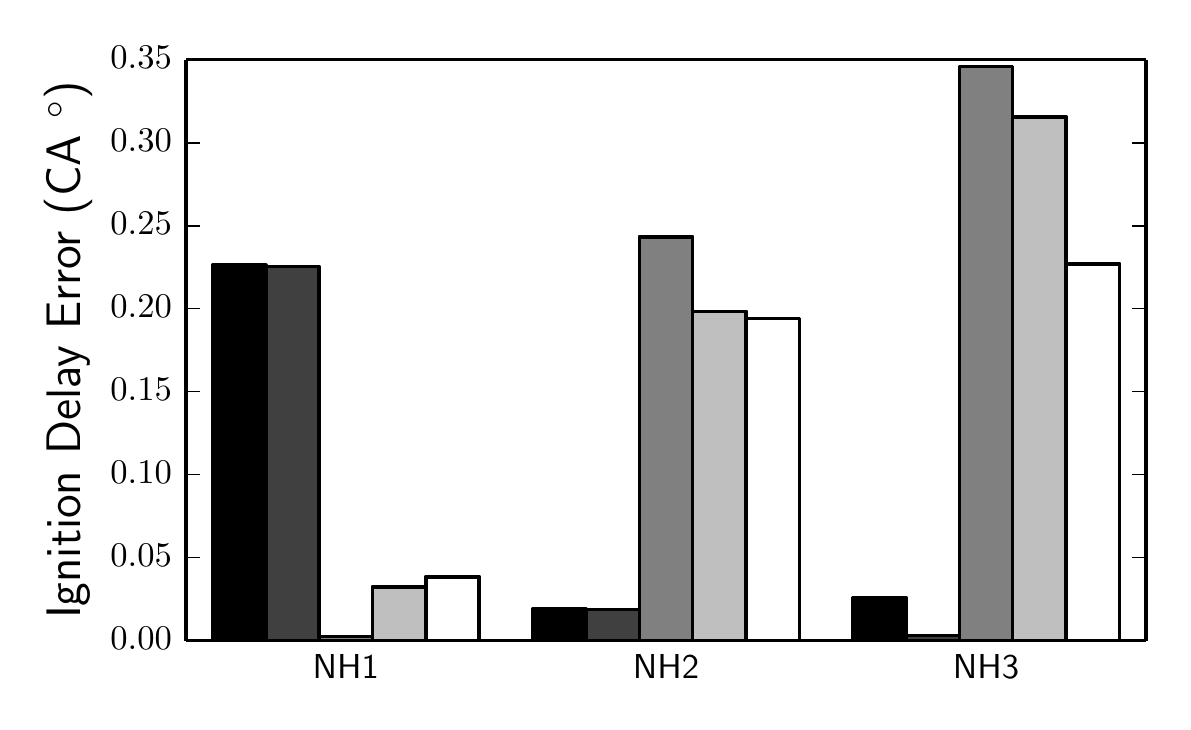}
	\includegraphics[width = 0.5\linewidth, keepaspectratio = true]{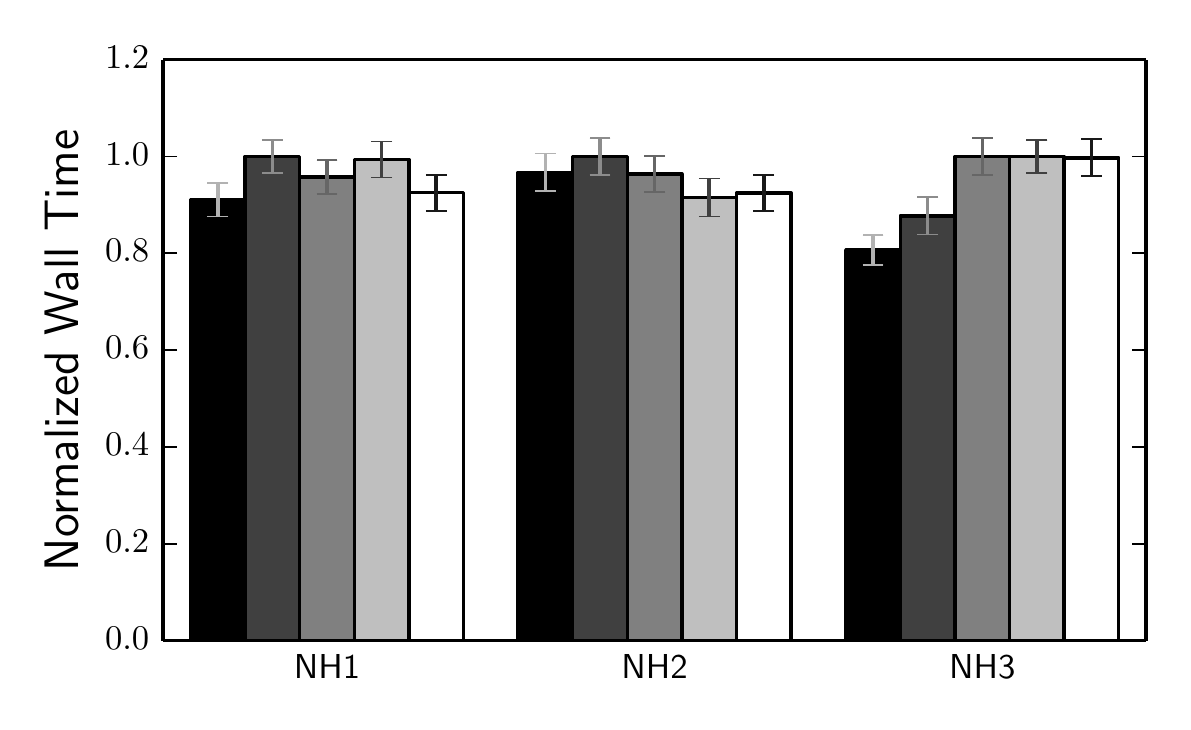}
}
\end{subcaptionbox} \\
\begin{subcaptionbox}{$\Edrgep = $ \num{e-3} \label{subfig:nhept-hcci-higherr}}
{
	\centering
	\includegraphics[width = 0.5\linewidth, keepaspectratio = true]{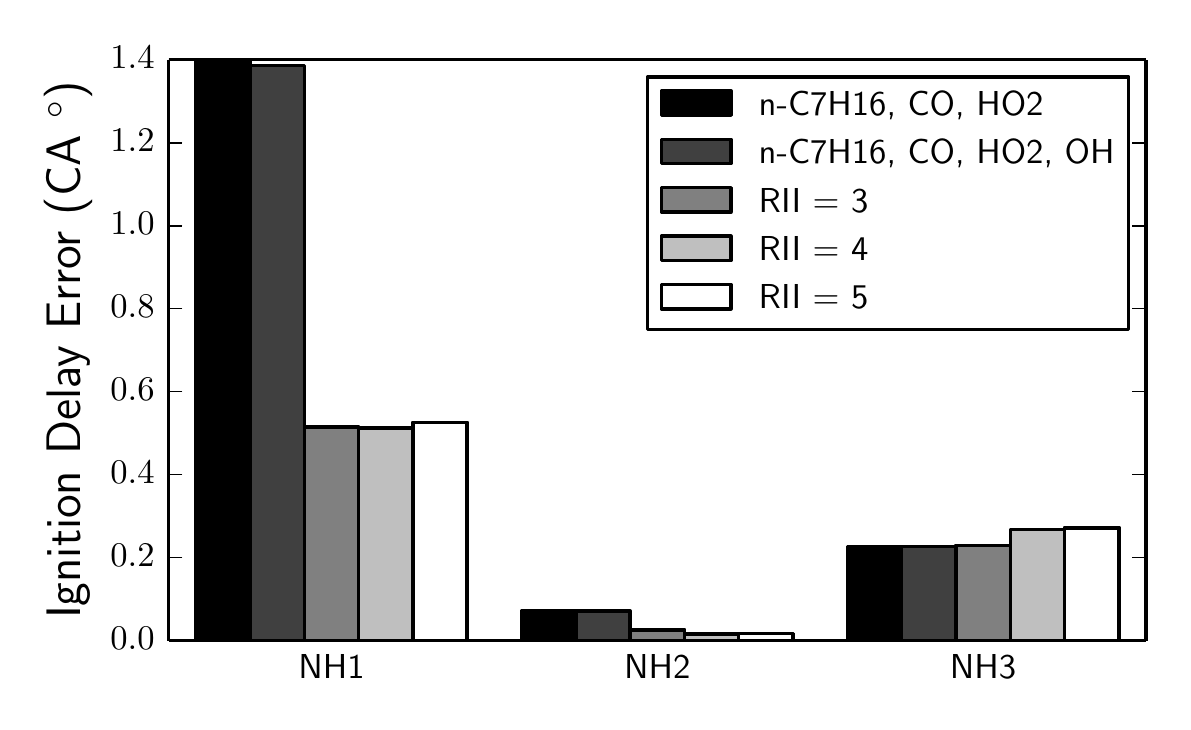}
	\includegraphics[width = 0.5\linewidth, keepaspectratio = true]{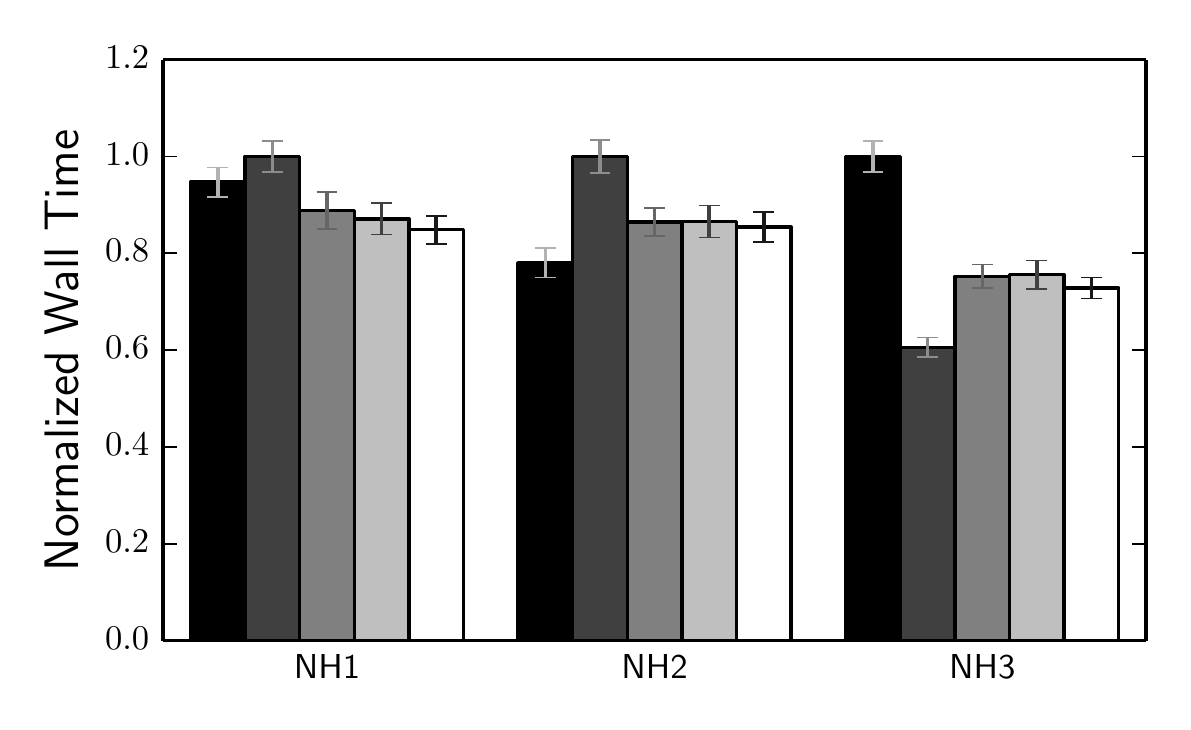}
}
\end{subcaptionbox}

\caption{Comparison of the ignition delay error (in crank angle degrees) and simulation wall time normalized by the longest time in each case for single-cell HCCI simulations of n-heptane at conditions listed in Table~\ref{tab:nhept_hcci_conditions}.}
\label{fig:n-hept hcci dynamic}
\end{figure}

Next, the performance and accuracy of the RII method will be compared to that of the two static target species sets in single-cell engine simulations at HCCI conditions adapted from Liang et al.~\cite{liang_dac} (Table~\ref{tab:nhept_hcci_conditions}).
Figure~\ref{fig:n-hept hcci dynamic} compares the accuracy and wall-clock time of the two static target species sets with three, four, and five RII target species criteria, as used in Section~\ref{sec:nhept-conv}.
The simulation wall-clock times shown are normalized by the longest run for each starting condition.

Figure~\ref{subfig:nhept-hcci-lowerr} compares the static target sets to the RII target species criteria, with a DRGEP threshold of $\Edrgep =$ \num{e-4}.
The RII target species criteria are more accurate for the low equivalence ratio case (NH1), but are less accurate for the larger equivalence ratio cases (NH2, NH3).  
Nevertheless, the maximum error for the static targets is \SI{0.23}{\degree}, while the maximum error for the RII target species criteria ranges from \SIrange{0.35}{0.23}{\degree} for three to five RII targets, respectively.
The speed of the RII simulations is roughly the same as the static target sets; although for the $\phi = 1.2$ case (NH3), the RII target species criteria are approximately \SIrange[range-phrase = --, range-units = single]{15}{20}{\percent} slower than the static target sets.

The inaccuracy of the RII target species criteria in cases NH2 and NH3 is the result of the relatively large mass fraction cutoff used (i.e., \num{e-8}).  This large mass fraction cutoff delays the selection of \ce{CH2CHO} and \ce{CH2CH2CHO} (early \ce{CO2} precursors) as target species, which in turn delays ignition.
By lowering the mass fraction cutoff (e.g., to \num{e-10}) the accuracy of the RII target species criteria becomes equivalent to that of the static target species sets; however, the computational efficiency of the RII method is decreased due to issues with biased RII values (as described in Section~\ref{sec:target selection}), and the static target species sets perform up to \SI{30}{\percent} faster for cases NH2 and NH3.
Although use of the larger mass fraction cutoff induces higher error in these cases, in most situations it provides a good balance between accuracy and efficiency.

Figure~\ref{subfig:nhept-hcci-higherr} compares the various target sets with a DRGEP threshold of $\Edrgep = $ \num{e-3}.
At this larger DRGEP threshold, the RII target species criteria are considerably more accurate than the static targets for case NH1, with a maximum error of $\sim$\SI{0.5}{\degree} compared to $\sim$\SI{1.4}{\degree} for the static target sets (both with and without \ce{OH}).
For cases NH2 and NH3, the accuracy of RII target species criteria are comparable to the static target sets.
The simulation wall-clock times show again that the speeds of RII and static target sets are roughly equivalent for this higher DRGEP threshold.
A speedup factor of approximately \SIrange[range-phrase = --, range-units = single]{4.5}{6}{$\times$} over the detailed mechanism was achieved for n-heptane with both the static target species sets and RII target species criteria.

\subsection{Isopentanol Mechanism}

Isopentanol is a next-generation biofuel with the potential to greatly reduce \ce{NO_x} and particulate emissions; as biofuels are produced from renewable sources, net \ce{CO2} emissions can be reduced as well \cite{iso-pentanol_mech}.
Less hygroscopic, corrosive, and miscible in water than ethanol~\cite{isopentanol_hcci, iso-pentanol_mech}, isopentanol is more compatible with the existing fuel infrastructure.
In addition, isopentanol has a volumetric energy density over \SI{30}{\percent} higher than that of ethanol~\cite{isopentanol_old_mech}.
These factors make isopentanol a promising alternative for future combustion devices, either by itself or blended with gasoline.

Isopentanol, a \ce{C5} alcohol with a methyl branch, exhibits largely different chemistry than a typical alkane fuel, including minimal NTC behavior and a single-stage ignition process.
As such, it is a good choice to demonstrate the general applicability of the RII method.
A detailed mechanism from Sarathy et al.~\cite{iso-pentanol_mech} with 360 species was used to study constant-volume autoignition problems and single-cell HCCI simulations.

\subsection{Isopentanol Constant Volume Simulations}

First, the accuracy of the RII method will be compared to that of static target species sets using constant-volume simulations with initial conditions of \SIlist{5;20}{\atm}, \SIrange{700}{1100}{\kelvin}, and $\phi = \set{0.38, 0.6, 1.0}$.
The static target sets again used a mass fraction cutoff of \num{e-30}, while the RII target species criteria used a cutoff of \num{e-8}.
For this mechanism, the highly connected species (as determined by the criteria detailed in Section~\ref{sec:target selection}) were \ce{CH3}, \ce{H}, \ce{HO2}, \ce{O2}, and \ce{OH}; the mass fraction cutoff for these species was relaxed to \num{e-15}.
For all cases, the DRGEP cutoff threshold $\Edrgep$ was set to \num{5e-3}.

Table~\ref{tab:ipent-conv-errorstats} compares the error statistics for the various target species sets.
It is seen that including \ce{OH} in the static target species set again increases the accuracy from a maximum error of \SI{12.3}{\percent} to \SI{9.51}{\percent}, while the RII target species criteria are even more accurate with a maximum error of \SIrange{6.86}{5.31}{\percent} for the three to five RII target species criteria, respectively.
The average percent error over all the cases is significantly lower for the RII target species criteria as well, ranging from \SIrange{2.83}{2.00}{\percent}, compared to \SIrange{4.43}{3.40}{\percent} for the static target sets.

\subsection{Isopentanol HCCI Engine Simulations}

Next, the RII method was tested with single-cell engine simulations at HCCI conditions listed in Table~\ref{tab:ipent_hcci_conditions}, adapted from those given by Yang et al.~\cite{isopentanol_hcci}.
The static target sets tested were the same as those used in the constant-volume autoignition simulations.

To achieve a simplified exhaust gas recirculation (EGR) process, the engine cycle was first simulated at the conditions listed.
At \SI{122}{\degree} ATDC (corresponding to exhaust valve opening), the state of the engine cell was extracted and assumed to be cooled and throttled to \SI{367}{\kelvin} and \SI{202}{\kilo\pascal} (i.e., the exit temperature and pressure of the EGR cooling cycle in \cite{isopentanol_hcci}), before being mixed adiabatically on a per mass basis with the inlet gas.
In addition, the amount of isopentanol was adjusted to keep the charge mass equivalence ratio,
\begin{equation}
\phi_m = \frac{(F/C)}{(F/A)_{stoich}} \;,
\end{equation}
constant following the procedure of Yang et al.~\cite{isopentanol_hcci}. $(F/C)$ is the mass ratio of fuel to total charge mass (i.e. fresh air and recycled exhaust gas) and $(F/A)_{stoich}$ is the stoichiometric fuel/air mass ratio.

\begin{figure}
\begin{subcaptionbox}{$\Edrgep = $ \num{e-3} \label{subfig:ipent-hcci-lowerr}}
{
	\centering
	\includegraphics[width = 0.5\linewidth, keepaspectratio = true]{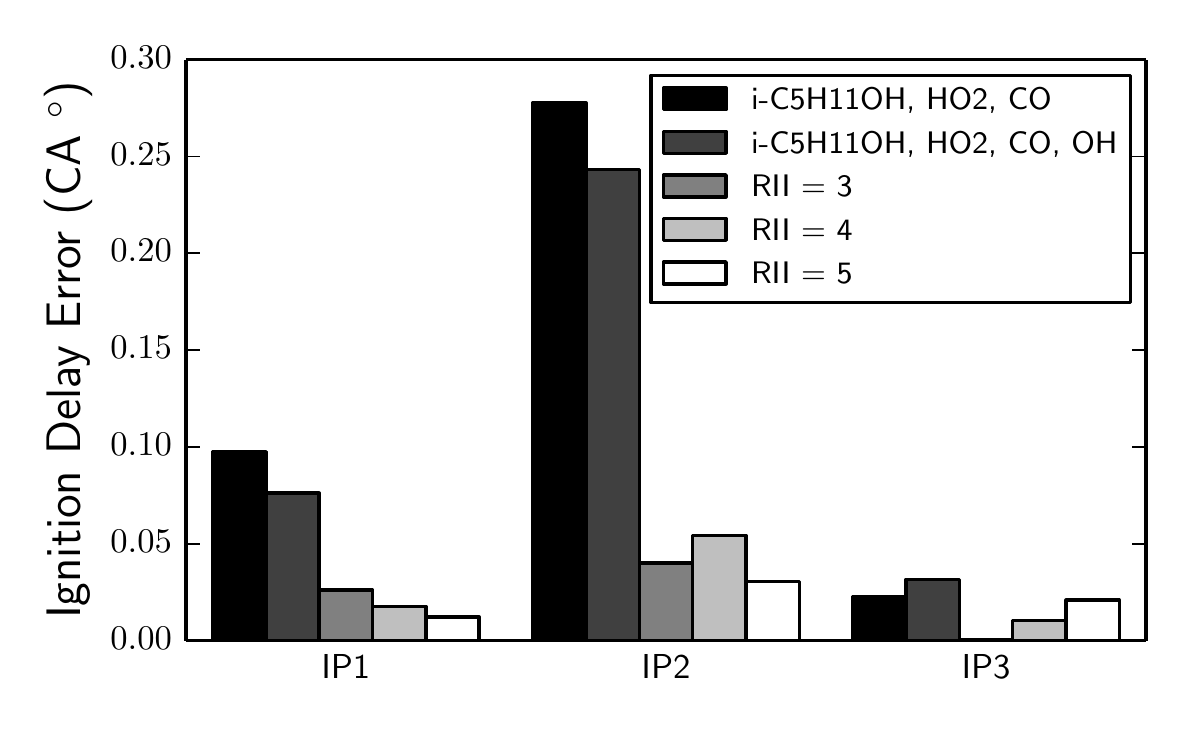}
	\includegraphics[width = 0.5\linewidth, keepaspectratio = true]{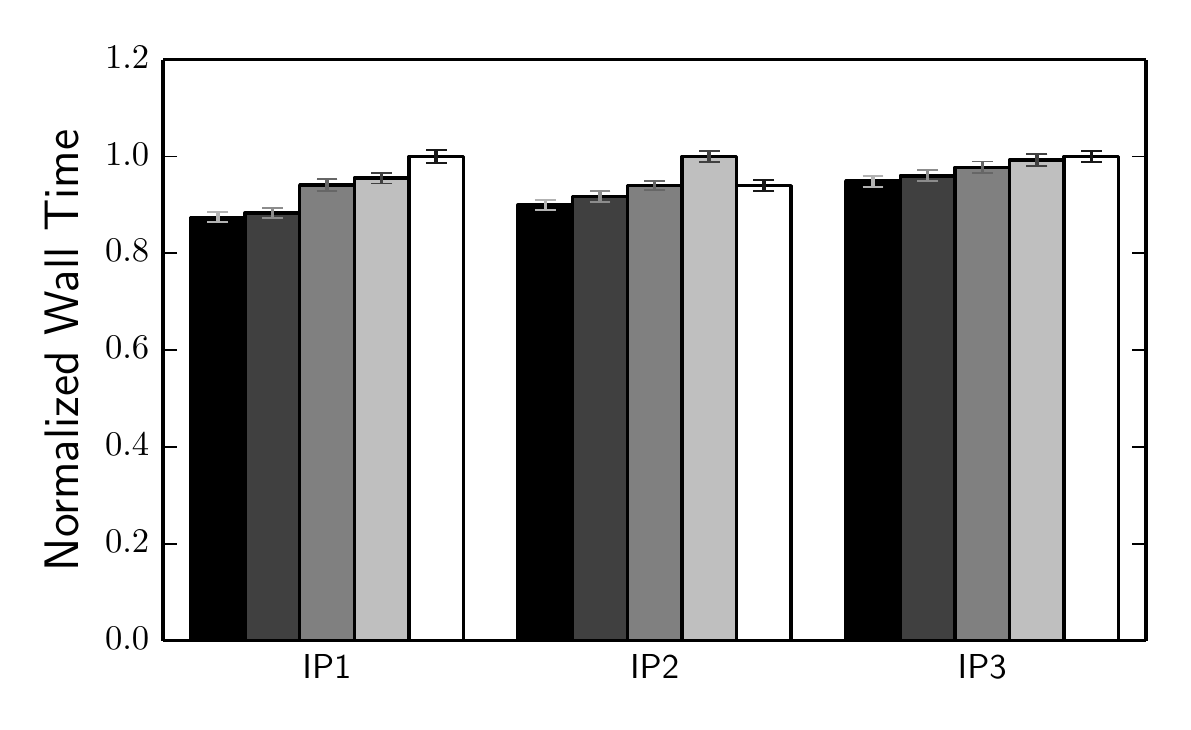}
}
\end{subcaptionbox} \\
\begin{subcaptionbox}{$\Edrgep = $ \num{5e-3} \label{subfig:ipent-hcci-higherr}}
{
	\centering
	\includegraphics[width = 0.5\linewidth, keepaspectratio = true]{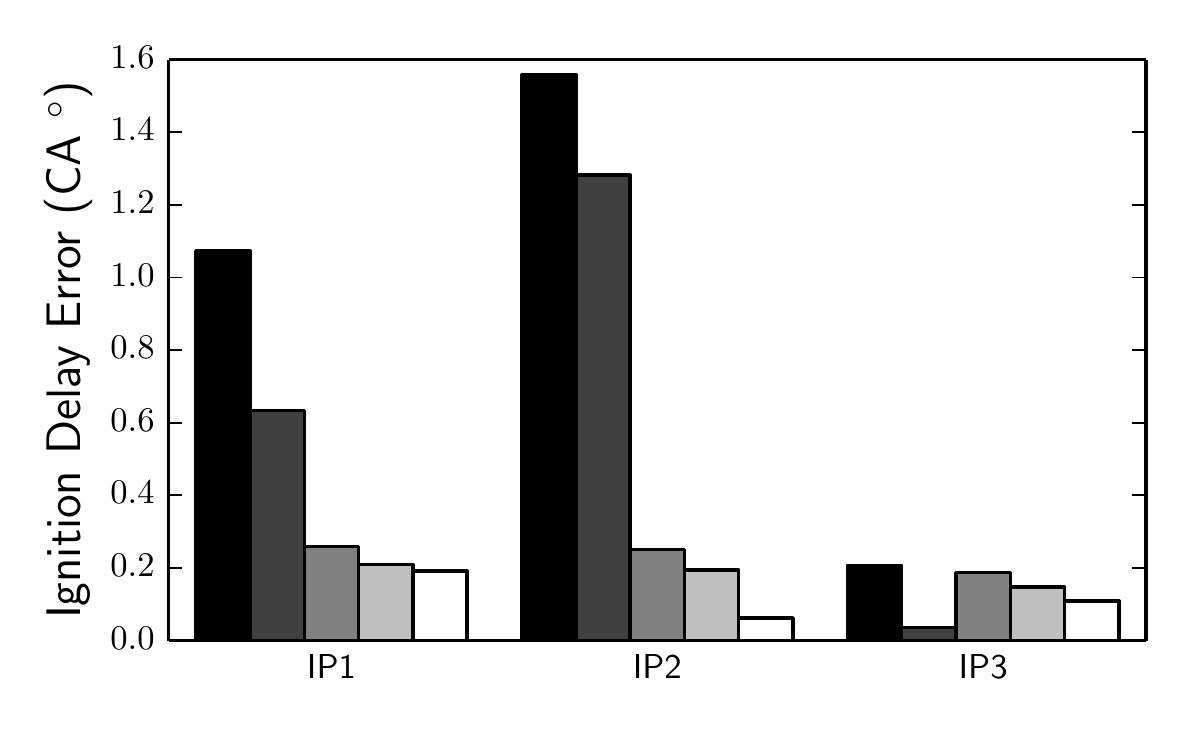}
	\includegraphics[width = 0.5\linewidth, keepaspectratio = true]{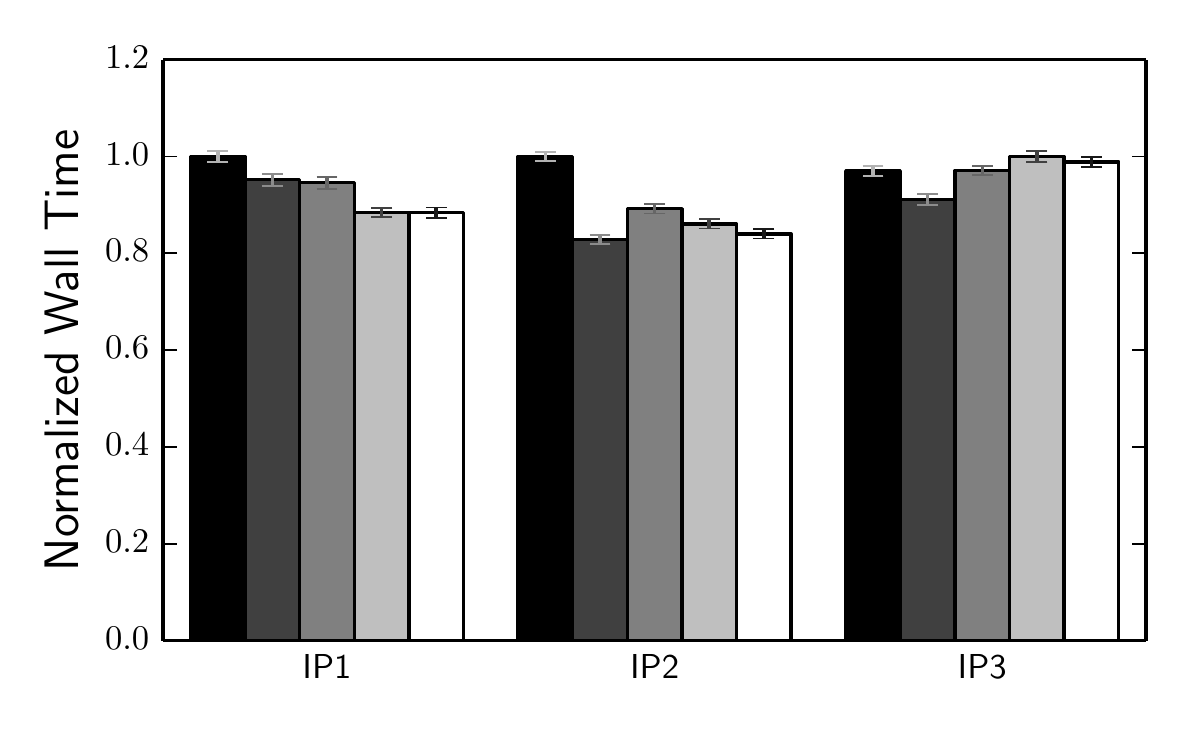}
}
\end{subcaptionbox}
\caption{Comparison of the ignition delay error (in crank angle degrees) and simulation wall time normalized by the longest time in each case for single-cell HCCI simulations of isopentanol at conditions in Table~\ref{tab:ipent_hcci_conditions}.}
\label{fig:i-pent hcci}
\end{figure}

Figure~\ref{fig:i-pent hcci} compares the errors in ignition delay and wall-clock times of the various target species sets; the latter were again normalized by the longest time in each case.
At the smaller DRGEP threshold (\num{e-3}), the static target species sets have a maximum error of \SIrange{0.24}{0.28}{\degree} with and without \ce{OH}, respectively, while the RII target species criteria have a maximum error of approximately \SI{0.05}{\degree} (Fig.~\ref{subfig:ipent-hcci-lowerr}).
The RII target species criteria performed around \SIrange{5}{10}{\percent} slower at this DRGEP threshold.

For a higher DRGEP threshold (\num{5e-3}), the RII targets species criteria perform more accurately than the static target species sets for almost all conditions shown in Fig.~\ref{subfig:ipent-hcci-higherr}.
The static target sets result in maximum errors of \SIrange{1.28}{1.56}{\degree} with and without \ce{OH}, respectively, while the RII target species criteria produce maximum errors of \SIrange{0.25}{0.19}{\degree} for three to five RII targets, respectively.
In terms of computational efficiency, the various target sets performed similarly at this DRGEP threshold, with the static target sets operating faster for some conditions and the RII target species criteria faster for others.
A speedup factor of approximately \SIrange[range-phrase = --, range-units = single]{2}{3}{$\times$} over the detailed mechanism was achieved for isopentanol with both the static target species sets and RII target species criteria.

\section{Conclusions and Future Work}
\label{sec:Conclusion}

The current methodology of DRGEP-based DAC simulations typically utilizes a static (unchanging) set of target species for the DRGEP method.
However, as demonstrated in Section~\ref{sec:Applicability}, while using a single static target species set (e.g.,\ \liangtargets{}) may work well for simple simulations, this approach could suffer in accuracy for different problems and\slash or combustion conditions.
In an attempt to improve accuracy of a static target species set in general, additional key species can be added (e.g.,\ \ce{OH}); however, this requires expert knowledge of the chemical mechanism at hand and may adversely affect performance.
Further, for a complex multidimensional simulation the range of combustion conditions encountered cannot be known with certainty a priori.
While multiple cases could be run to test different target species sets, the prohibitive computational expense of this operation makes it difficult to ensure the applicability of a selected target species set.

In this work, a novel method of automatically determining appropriate target species for DRGEP-based DAC simulations was developed and implemented that relies solely on the local thermo-chemical state: the relative importance index (RII) method.
As shown in Section~\ref{sec:target selection}, the RII method selects target species based on their relevance to the local chemistry.
For single-cell HCCI engine simulations of n-heptane and isopentanol, the RII method was demonstrated to match the accuracy of conventional static target species sets at small DRGEP thresholds, while maintaining higher accuracy at larger DRGEP thresholds.
In all cases, RII performed similarly in terms of execution speed to the static targets sets.
In addition, the same RII target species criteria were shown to be more accurate than static target sets in the more stringent cases of constant-volume autoignition studies for both the n-heptane and isopentanol mechanisms.
These factors make the RII method a promising candidate to maintain accuracy and ensure efficient performance for complex multidimensional simulations.
Additionally, automating the target selection process with RII greatly reduces the user knowledge required to select appropriate DRGEP target species, simplifying the use of a DRGEP-based DAC method for new problem types and reaction mechanisms.

Further improvement to the RII method is likely to be achieved through the following avenues.
First, the RII bias filtering method discussed in Section~\ref{sec:target selection} is largely empirical, and while sufficient in the context of this work, a more rigorous method may improve the accuracy of the RII method.
Second, a natural extension for the RII method is the automatic selection of a DRGEP threshold derived from the local thermo-chemical state.
In the DAC-based simulation efforts demonstrated thus far, the DRGEP threshold value is typically a static value selected based on past experience, which provides only rudimentary a priori error control.
As a result, the DRGEP threshold is often set to a conservative value in order to ensure accuracy; however, as shown by Shi et al.~\cite{shi_dac}, the DRGEP threshold can be increased in certain cases---e.g.,\ after combustion has completed---with minimal effect on the accuracy of the simulation while improving execution speed.
Gou et al.~\cite{dac_gou} proposed an error controlled DAC scheme with some success; however, it is noted that their method relied on data tabulated from homogeneous ignition studies and therefore may not be applicable for general multidimensional simulations.
Therefore, more investigation is needed in this direction.
Finally, although the accuracy of the DRG method is less sensitive to the proper selection of target species, there are cases where DRG-based DAC schemes would benefit from the use of an automated target selection method, e.g. by removing the fuel molecule from the target species set after ignition or in situations where combustion is dominated by a single species not included in the static target species set.
To improve the generality of the RII method, its applicability to DRG-based DAC schemes should be thoroughly investigated.

\section*{Acknowledgments}

This work was supported by the Combustion Energy Frontier Research Center, an Energy Frontier Research Center funded by the US Department of Energy, Office of Science, Office of Basic Energy Sciences under award number DE-SC0001198, and the National Science Foundation Graduate STEM Fellows in K-12 Education program.

\pagebreak
\begin{appendices}
\section*{Appendix: Simple Example of RII Biasing}
\begin{figure}
\centering
\begin{subfigure}[t]{\linewidth}
{
\includegraphics[width = \linewidth, keepaspectratio = true]{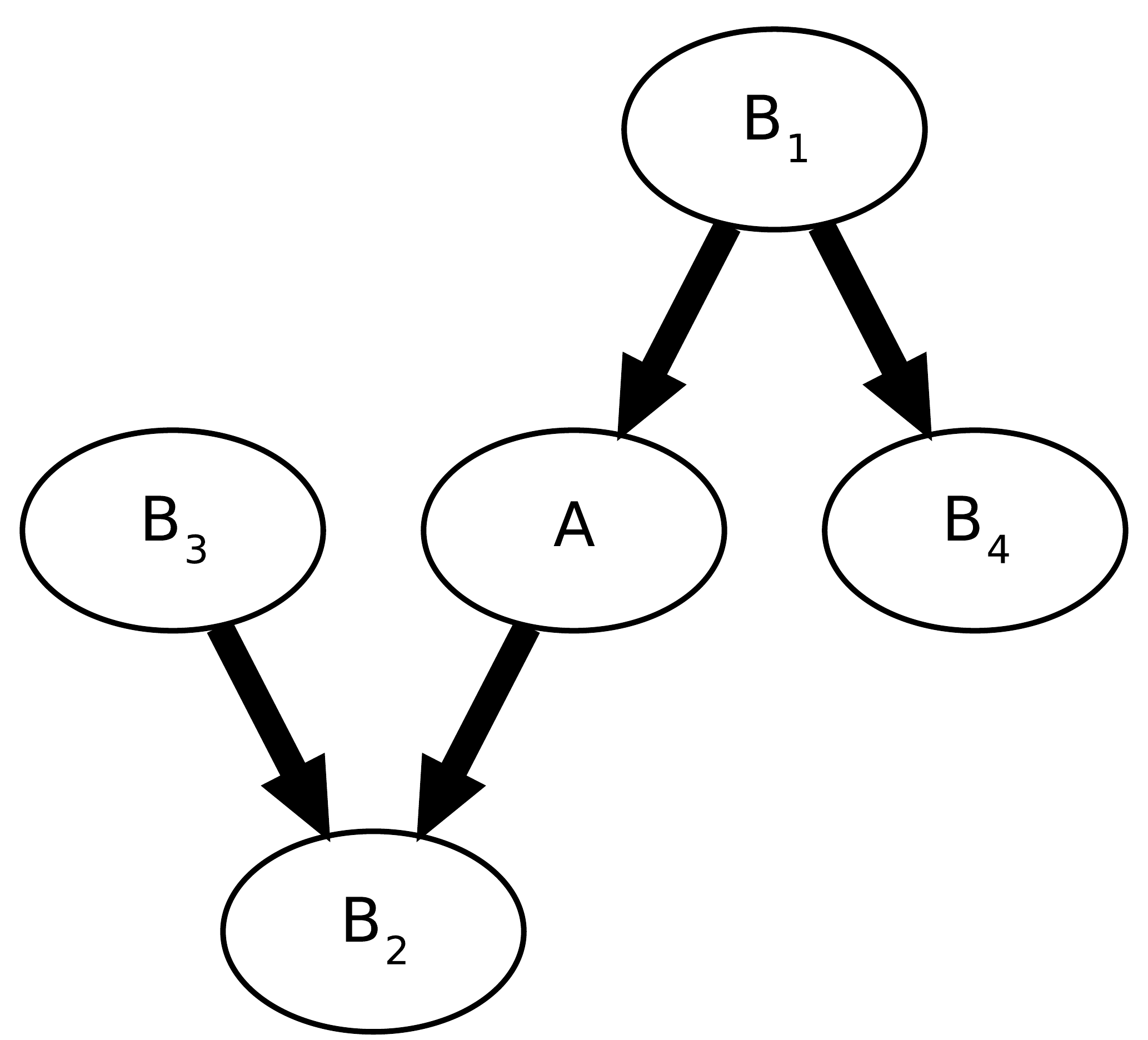}
}
\end{subfigure}
\caption{A simple example of reaction pathways that may lead to RII biasing.  Species $B_1$--$B_4$ participate in reactions with $A$, and hence are its direct neighbors.}
\label{fig:simpleex}
\end{figure}

Consider the reaction pathways presented in Fig.~\ref{fig:simpleex}, which gives an example of two steps in a long chain of large hydrocarbon breakdown reactions.
Species $A$ participates in only two reactions: formation through the decomposition of $B_1$ (forming $A$ and $B_4$), and a reaction with $B_3$ forming $B_2$.  
Note that the neighboring species of $A$ in this simple example are $B_1$--$B_4$.
It will be demonstrated that the RII of a species $A$ in this situation can be large enough that the species may be selected as a target species even if $A$ is only weakly reacting.
Hence for a species with only a few neighbors, an additional criterion to determine the importance of a species is needed to supplement the RII.

If the rate of production of $A$ ($P_A$) is greater than the rate of consumption of $A$ ($C_A$), from Eq.~\eqref{eq:DIC} the DICs $\DICeq{AB_1}$ and $\DICeq{AB_4}$ must be unity as $P_A > C_A$.  
In this case, the DICs $\DICeq{AB_2}$ and $\DICeq{AB_3}$ will simply be equal to the ratio of consumption to production of $A$, $\frac{C_A}{P_A}$.
Therefore:
\begin{equation}
\text{Row Average(A)} = \frac{2 + 2 \times \frac{C_A}{P_A}}{4}.
\end{equation}
If on the other hand, $A$ is being consumed faster than it is produced it follows that:
\begin{equation}
\text{Row Average(A)} = \frac{2 + 2 \times \frac{P_A}{C_A}}{4}.
\end{equation}
From which it is seen the row average of $A$ is bounded between \num{0.5} and \num{1.0}.
Even if the production and consumption of $A$ is highly unbalanced, the lower bound on the row average of $A$ is very large compared to those shown in Fig.~\ref{fig:riirow}.

Further, if $B_1$--$B_4$ react in relatively few reactions, as will typically be the case for a long chain of large hydrocarbon breakdown reactions, the column sum of $A$ will naturally tend to be large.  
This phenomenon is analogous to the discussion in Section~\ref{sec:riirow}; as each neighboring species $B_i$ has few neighbors, $A$ will likely be involved in a large portion of the production or consumption of $B_i$.
Hence the DICs making up the column sum of $A$, $\DICeq{B_1A}$, $\DICeq{B_2A}$, $\DICeq{B_3A}$, and $\DICeq{B_4A}$, will tend to be closer to \num{1.0} in value.  
This implies that although the \text{Column Sum(A)} is bounded between \num{0.0} and \num{4.0}, it may be closer to \num{4.0}.

Therefore, the \text{RII(A)} can easily be larger than the RII's of commonly used target species presented in Fig.~\ref{fig:rii}.
Consider, if we simply take the row average and column sums to be the midpoints of their bounds (i.e., $\text{Row Average(A)} = \num{0.75}$, and $\text{Column Sum(A)} = \num{2.0}$), the \text{RII(A)} will be equal to \num{1.5}, comparable to that of \ce{OH} and greater than that of \ce{HO2} in the sample ignition cases (Fig.~\ref{fig:rii}).
Finally it is noted that the above discussion applies whether $A$ is strongly reacting or not; hence, in such a situation the \text{RII(A)} may not accurately reflect the true importance of $A$.

Such behavior is only possible for a species that participates in just a handful of reactions, with only a few neighbors.
For a species with more neighbors, the RII will better reflect the relative activity of the species as seen in Section~\ref{sec:rii_def}, because it becomes unlikely that the production and consumption rates of the species and its neighbors will exhibit the strongly coupled behavior demonstrated in this example. 
Further, since the species $A$ in this example is only a single stage in a long breakdown chain, it is not expected to exist in high concentrations for long periods.
Instead it is more likely that species $A$ will be present in large quantities only when it is strongly reacting; hence using a relatively large mass fraction cutoff (e.g., \num{e-8}) will tend to eliminate species $A$ from consideration as a target species when it is only weakly reacting.

\end{appendices}

\pagebreak
\section*{References}
\bibliography{Bibliography}

\begin{thebibliography}{56}
\expandafter\ifx\csname natexlab\endcsname\relax\def\natexlab#1{#1}\fi
\providecommand{\bibinfo}[2]{#2}

\bibitem[{Correa(1998)}]{correa_turbines}
\bibinfo{author}{S.~M. Correa}, \bibinfo{journal}{Proc. Combust. Inst.}
  \bibinfo{volume}{27} (\bibinfo{year}{1998}) \bibinfo{pages}{1793--1807}.

\bibitem[{Lu and Law(2009)}]{lu_towardrealistic}
\bibinfo{author}{T.~Lu}, \bibinfo{author}{C.~K. Law}, \bibinfo{journal}{Prog.
  Energy Combust. Sci.} \bibinfo{volume}{35} (\bibinfo{year}{2009})
  \bibinfo{pages}{192--215}.

\bibitem[{Shanbhogue et~al.(2009)Shanbhogue, Husain, and
  Lieuwen}]{shan_leanblowout}
\bibinfo{author}{S.~J. Shanbhogue}, \bibinfo{author}{S.~Husain},
  \bibinfo{author}{T.~Lieuwen}, \bibinfo{journal}{Prog. Energy Combust. Sci.}
  \bibinfo{volume}{35} (\bibinfo{year}{2009}) \bibinfo{pages}{98--120}.

\bibitem[{Yao et~al.(2009)Yao, Zheng, and Liu}]{yao_hcci}
\bibinfo{author}{M.~Yao}, \bibinfo{author}{Z.~Zheng}, \bibinfo{author}{H.~Liu},
  \bibinfo{journal}{Prog. Energy Combust. Sci.} \bibinfo{volume}{35}
  (\bibinfo{year}{2009}) \bibinfo{pages}{398--437}.

\bibitem[{Tonse et~al.(2003)Tonse, Moriarty, Frenklach, and Brown}]{tonse_cfd}
\bibinfo{author}{S.~R. Tonse}, \bibinfo{author}{N.~W. Moriarty},
  \bibinfo{author}{M.~Frenklach}, \bibinfo{author}{N.~J. Brown},
  \bibinfo{journal}{Int. J. Chem. Kinet.} \bibinfo{volume}{35}
  (\bibinfo{year}{2003}) \bibinfo{pages}{438--452}.

\bibitem[{Liang et~al.(2007)Liang, Kong, Jung, and Reitz}]{liang_semiimplicit}
\bibinfo{author}{L.~Liang}, \bibinfo{author}{S.-C. Kong},
  \bibinfo{author}{C.~Jung}, \bibinfo{author}{R.~D. Reitz},
  \bibinfo{journal}{J. Eng. Gas Turb. Power} \bibinfo{volume}{129}
  (\bibinfo{year}{2007}) \bibinfo{pages}{271--278}.

\bibitem[{Shi et~al.(2012)Shi, Green, Wong, and Oluwole}]{shi_hybrid}
\bibinfo{author}{Y.~Shi}, \bibinfo{author}{W.~H. Green}, \bibinfo{author}{H.-W.
  Wong}, \bibinfo{author}{O.~O. Oluwole}, \bibinfo{journal}{Combust. Flame}
  \bibinfo{volume}{159} (\bibinfo{year}{2012}) \bibinfo{pages}{2388--2397}.

\bibitem[{Shi et~al.(2010)Shi, Liang, Ge, and Reitz}]{shi_dac}
\bibinfo{author}{Y.~Shi}, \bibinfo{author}{L.~Liang}, \bibinfo{author}{H.-W.
  Ge}, \bibinfo{author}{R.~D. Reitz}, \bibinfo{journal}{Combust. Theory Model.}
  \bibinfo{volume}{14} (\bibinfo{year}{2010}) \bibinfo{pages}{69--89}.

\bibitem[{Lu and Law(2005)}]{drg_lu}
\bibinfo{author}{T.~Lu}, \bibinfo{author}{C.~K. Law}, \bibinfo{journal}{Proc.
  Combust. Inst.} \bibinfo{volume}{30} (\bibinfo{year}{2005})
  \bibinfo{pages}{1333--1341}.

\bibitem[{Lu and Law(2006)}]{drg_lu2}
\bibinfo{author}{T.~Lu}, \bibinfo{author}{C.~K. Law},
  \bibinfo{journal}{Combust. Flame} \bibinfo{volume}{144}
  (\bibinfo{year}{2006}) \bibinfo{pages}{24--36}.

\bibitem[{Lu and Law(2008)}]{stategies_lu}
\bibinfo{author}{T.~Lu}, \bibinfo{author}{C.~K. Law},
  \bibinfo{journal}{Combust. Flame} \bibinfo{volume}{154}
  (\bibinfo{year}{2008}) \bibinfo{pages}{153--163}.

\bibitem[{Pepiot-Desjardins and Pitsch(2008)}]{drgep_pepiot}
\bibinfo{author}{P.~Pepiot-Desjardins}, \bibinfo{author}{H.~Pitsch},
  \bibinfo{journal}{Combust. Flame} \bibinfo{volume}{154}
  (\bibinfo{year}{2008}) \bibinfo{pages}{67--81}.

\bibitem[{Rabitz et~al.(1983)Rabitz, Kramer, and Dacol}]{rabitz_1983}
\bibinfo{author}{H.~Rabitz}, \bibinfo{author}{M.~Kramer},
  \bibinfo{author}{D.~Dacol}, \bibinfo{journal}{Annu. Rev. Phys. Chem.}
  \bibinfo{volume}{34} (\bibinfo{year}{1983}) \bibinfo{pages}{419--461}.

\bibitem[{Tur{\'a}nyi(1990{\natexlab{a}})}]{turanyi_sa_1}
\bibinfo{author}{T.~Tur{\'a}nyi}, \bibinfo{journal}{New J. Chem.}
  \bibinfo{volume}{14} (\bibinfo{year}{1990}{\natexlab{a}})
  \bibinfo{pages}{795--803}.

\bibitem[{Tur{\'a}nyi(1990{\natexlab{b}})}]{turanyi_sa_2}
\bibinfo{author}{T.~Tur{\'a}nyi}, \bibinfo{journal}{J. Math. Chem.}
  \bibinfo{volume}{5} (\bibinfo{year}{1990}{\natexlab{b}})
  \bibinfo{pages}{203--248}.

\bibitem[{Vajda et~al.(1985)Vajda, Valko, and Tur{\'a}nyi}]{vajda_pca}
\bibinfo{author}{S.~Vajda}, \bibinfo{author}{P.~Valko},
  \bibinfo{author}{T.~Tur{\'a}nyi}, \bibinfo{journal}{Int. J. Chem. Kinet.}
  \bibinfo{volume}{17} (\bibinfo{year}{1985}) \bibinfo{pages}{55--81}.

\bibitem[{L{\o}v{\aa}s et~al.(2000)L{\o}v{\aa}s, Nilsson, and
  Mauss}]{lovas_loi}
\bibinfo{author}{T.~L{\o}v{\aa}s}, \bibinfo{author}{D.~Nilsson},
  \bibinfo{author}{F.~Mauss}, \bibinfo{journal}{Proc. Combust. Inst.}
  \bibinfo{volume}{28} (\bibinfo{year}{2000}) \bibinfo{pages}{1809--1815}.

\bibitem[{L{\o}v{\aa}s et~al.(2002)L{\o}v{\aa}s, Mauss, Hasse, and
  Peters}]{lovas_loi2}
\bibinfo{author}{T.~L{\o}v{\aa}s}, \bibinfo{author}{F.~Mauss},
  \bibinfo{author}{C.~Hasse}, \bibinfo{author}{N.~Peters},
  \bibinfo{journal}{Proc. Combust. Inst.} \bibinfo{volume}{29}
  (\bibinfo{year}{2002}) \bibinfo{pages}{1403--1410}.

\bibitem[{L{\o}v{\aa}s(2009)}]{lovas_2009}
\bibinfo{author}{T.~L{\o}v{\aa}s}, \bibinfo{journal}{Combust. Flame}
  \bibinfo{volume}{156} (\bibinfo{year}{2009}) \bibinfo{pages}{1348--1358}.

\bibitem[{Lam and Coussis(1989)}]{csp1}
\bibinfo{author}{S.~Lam}, \bibinfo{author}{D.~Coussis}, \bibinfo{journal}{Proc.
  Combust. Inst.} \bibinfo{volume}{22} (\bibinfo{year}{1989})
  \bibinfo{pages}{931--941}.

\bibitem[{Lam(1993)}]{csp2}
\bibinfo{author}{S.~Lam}, \bibinfo{journal}{Combust. Sci. Tech.}
  \bibinfo{volume}{89} (\bibinfo{year}{1993}) \bibinfo{pages}{375--404}.

\bibitem[{Lam and Goussis(1994)}]{csp3}
\bibinfo{author}{S.~Lam}, \bibinfo{author}{D.~Goussis}, \bibinfo{journal}{Int.
  J. Chem. Kinet.} \bibinfo{volume}{26} (\bibinfo{year}{1994})
  \bibinfo{pages}{461--486}.

\bibitem[{Valorani et~al.(2006)Valorani, Creta, Goussis, Lee, and
  Najm}]{valorani_csp}
\bibinfo{author}{M.~Valorani}, \bibinfo{author}{F.~Creta},
  \bibinfo{author}{D.~A. Goussis}, \bibinfo{author}{J.~C. Lee},
  \bibinfo{author}{H.~N. Najm}, \bibinfo{journal}{Combust. Flame}
  \bibinfo{volume}{146} (\bibinfo{year}{2006}) \bibinfo{pages}{29--51}.

\bibitem[{Valorani et~al.(2007)Valorani, Creta, Donato, Najm, and
  Goussis}]{valorani_csp2}
\bibinfo{author}{M.~Valorani}, \bibinfo{author}{F.~Creta},
  \bibinfo{author}{F.~Donato}, \bibinfo{author}{H.~N. Najm},
  \bibinfo{author}{D.~A. Goussis}, \bibinfo{journal}{Proc. Combust. Inst.}
  \bibinfo{volume}{31} (\bibinfo{year}{2007}) \bibinfo{pages}{483--490}.

\bibitem[{Niemeyer et~al.(2010)Niemeyer, Sung, and Raju}]{niemeyer_drgepsa}
\bibinfo{author}{K.~E. Niemeyer}, \bibinfo{author}{C.-J. Sung},
  \bibinfo{author}{M.~P. Raju}, \bibinfo{journal}{Combust. Flame}
  \bibinfo{volume}{157} (\bibinfo{year}{2010}) \bibinfo{pages}{1760--1770}.

\bibitem[{Niemeyer and Sung(2014)}]{Niemeyer:2014}
\bibinfo{author}{K.~E. Niemeyer}, \bibinfo{author}{C.-J. Sung},
  \bibinfo{journal}{Combust. Flame} \bibinfo{volume}{161}
  (\bibinfo{year}{2014}) \bibinfo{pages}{2752--2764}.

\bibitem[{Chapman and Underhill(1913)}]{qssa}
\bibinfo{author}{D.~L. Chapman}, \bibinfo{author}{L.~K. Underhill},
  \bibinfo{journal}{J. Chem. Soc., Trans.} \bibinfo{volume}{103}
  (\bibinfo{year}{1913}) \bibinfo{pages}{496--508}.

\bibitem[{Benson(1952)}]{pe_approx1}
\bibinfo{author}{S.~W. Benson}, \bibinfo{journal}{J. Chem. Phys.}
  \bibinfo{volume}{20} (\bibinfo{year}{1952}) \bibinfo{pages}{1605--1612}.

\bibitem[{Ramshaw(1980)}]{pe_approx2}
\bibinfo{author}{J.~D. Ramshaw}, \bibinfo{journal}{Phys. Fluids}
  \bibinfo{volume}{23} (\bibinfo{year}{1980}) \bibinfo{pages}{675}.

\bibitem[{Maas and Pope(1992)}]{ildm}
\bibinfo{author}{U.~Maas}, \bibinfo{author}{S.~B. Pope},
  \bibinfo{journal}{Combust. Flame} \bibinfo{volume}{88} (\bibinfo{year}{1992})
  \bibinfo{pages}{239--264}.

\bibitem[{Pope(1997)}]{pope1997computationally}
\bibinfo{author}{S.~Pope}, \bibinfo{journal}{Combust. Theory Model.}
  \bibinfo{volume}{1} (\bibinfo{year}{1997}) \bibinfo{pages}{41--63}.

\bibitem[{Tonse et~al.(1999)Tonse, Moriarty, Brown, and Frenklach}]{prism}
\bibinfo{author}{S.~R. Tonse}, \bibinfo{author}{N.~W. Moriarty},
  \bibinfo{author}{N.~J. Brown}, \bibinfo{author}{M.~Frenklach},
  \bibinfo{journal}{Israel J. Chem.} \bibinfo{volume}{39}
  (\bibinfo{year}{1999}) \bibinfo{pages}{97--106}.

\bibitem[{Christo et~al.(1996)Christo, Masri, Nebot, and Pope}]{Christo1996}
\bibinfo{author}{F.~Christo}, \bibinfo{author}{A.~Masri},
  \bibinfo{author}{E.~Nebot}, \bibinfo{author}{S.~Pope},
  \bibinfo{journal}{Proc. Combust. Inst.} \bibinfo{volume}{26}
  (\bibinfo{year}{1996}) \bibinfo{pages}{43--48}.

\bibitem[{Liang et~al.(2009{\natexlab{a}})Liang, Stevens, and
  Farrell}]{liang_dac}
\bibinfo{author}{L.~Liang}, \bibinfo{author}{J.~G. Stevens},
  \bibinfo{author}{J.~T. Farrell}, \bibinfo{journal}{Proc. Combust. Inst.}
  \bibinfo{volume}{32} (\bibinfo{year}{2009}{\natexlab{a}})
  \bibinfo{pages}{527--534}.

\bibitem[{Liang et~al.(2009{\natexlab{b}})Liang, Stevens, Raman, and
  Farrell}]{liang_dac_gas}
\bibinfo{author}{L.~Liang}, \bibinfo{author}{J.~G. Stevens},
  \bibinfo{author}{S.~Raman}, \bibinfo{author}{J.~T. Farrell},
  \bibinfo{journal}{Combust. Flame} \bibinfo{volume}{156}
  (\bibinfo{year}{2009}{\natexlab{b}}) \bibinfo{pages}{1493--1502}.

\bibitem[{He et~al.(2010)He, Androulakis, and Ierapetritou}]{he_element}
\bibinfo{author}{K.~He}, \bibinfo{author}{I.~P. Androulakis},
  \bibinfo{author}{M.~G. Ierapetritou}, \bibinfo{journal}{Chem. Eng. Sci.}
  \bibinfo{volume}{65} (\bibinfo{year}{2010}) \bibinfo{pages}{1173--1184}.

\bibitem[{Yang et~al.(2012)Yang, Ren, Lu, and Goldin}]{yang_dac_drg}
\bibinfo{author}{H.~Yang}, \bibinfo{author}{Z.~Ren}, \bibinfo{author}{T.~Lu},
  \bibinfo{author}{G.~M. Goldin}, \bibinfo{journal}{Combust. Theory Model.}
  \bibinfo{volume}{17} (\bibinfo{year}{2012}) \bibinfo{pages}{167--183}.

\bibitem[{Tosatto et~al.(2013)Tosatto, Bennett, and Smooke}]{tosatto_drg}
\bibinfo{author}{L.~Tosatto}, \bibinfo{author}{B.~A.~V. Bennett},
  \bibinfo{author}{M.~D. Smooke}, \bibinfo{journal}{Combust. Flame}
  \bibinfo{volume}{160} (\bibinfo{year}{2013}) \bibinfo{pages}{1572--1582}.

\bibitem[{Gou et~al.(2013)Gou, Chen, Sun, and Ju}]{dac_gou}
\bibinfo{author}{X.~Gou}, \bibinfo{author}{Z.~Chen}, \bibinfo{author}{W.~Sun},
  \bibinfo{author}{Y.~Ju}, \bibinfo{journal}{Combust. Flame}
  \bibinfo{volume}{160} (\bibinfo{year}{2013}) \bibinfo{pages}{225--231}.

\bibitem[{Sun et~al.(2010)Sun, Chen, Gou, and Ju}]{sun_2010}
\bibinfo{author}{W.~Sun}, \bibinfo{author}{Z.~Chen}, \bibinfo{author}{X.~Gou},
  \bibinfo{author}{Y.~Ju}, \bibinfo{journal}{Combust. Flame}
  \bibinfo{volume}{157} (\bibinfo{year}{2010}) \bibinfo{pages}{1298--1307}.

\bibitem[{Contino et~al.(2011)Contino, Jeanmart, Lucchini, and
  D'Errico}]{Contino2011}
\bibinfo{author}{F.~Contino}, \bibinfo{author}{H.~Jeanmart},
  \bibinfo{author}{T.~Lucchini}, \bibinfo{author}{G.~D'Errico},
  \bibinfo{journal}{Proc. Combust. Inst.} \bibinfo{volume}{33}
  (\bibinfo{year}{2011}) \bibinfo{pages}{3057 -- 3064}.

\bibitem[{Contino et~al.(2012)Contino, Lucchini, D'Errico, Duynslaegher, Dias,
  and Jeanmart}]{contino2012simulations}
\bibinfo{author}{F.~Contino}, \bibinfo{author}{T.~Lucchini},
  \bibinfo{author}{G.~D'Errico}, \bibinfo{author}{C.~Duynslaegher},
  \bibinfo{author}{V.~Dias}, \bibinfo{author}{H.~Jeanmart},
  \bibinfo{title}{Simulations of advanced combustion modes using detailed
  chemistry combined with tabulation and mechanism reduction techniques},
  \bibinfo{type}{Technical Report}, SAE Technical Paper 2012-01-0145,
  \bibinfo{year}{2012}.

\bibitem[{Ren et~al.(2014{\natexlab{a}})Ren, Liu, Lu, Lu, Oluwole, and
  Goldin}]{ren_2014}
\bibinfo{author}{Z.~Ren}, \bibinfo{author}{Y.~Liu}, \bibinfo{author}{T.~Lu},
  \bibinfo{author}{L.~Lu}, \bibinfo{author}{O.~O. Oluwole},
  \bibinfo{author}{G.~M. Goldin}, \bibinfo{journal}{Combust. Flame}
  \bibinfo{volume}{161} (\bibinfo{year}{2014}{\natexlab{a}})
  \bibinfo{pages}{127--137}.

\bibitem[{Ren et~al.(2014{\natexlab{b}})Ren, Xu, Lu, and Singer}]{ren_2014b}
\bibinfo{author}{Z.~Ren}, \bibinfo{author}{C.~Xu}, \bibinfo{author}{T.~Lu},
  \bibinfo{author}{M.~A. Singer}, \bibinfo{journal}{J. Comput. Phys.}
  \bibinfo{volume}{263} (\bibinfo{year}{2014}{\natexlab{b}})
  \bibinfo{pages}{19--36}.

\bibitem[{Lu and Law(2006)}]{lu_2006}
\bibinfo{author}{T.~Lu}, \bibinfo{author}{C.~K. Law},
  \bibinfo{journal}{Combust. Flame} \bibinfo{volume}{146}
  (\bibinfo{year}{2006}) \bibinfo{pages}{472--483}.

\bibitem[{Niemeyer and Sung(2011)}]{niemeyer_graphsearch}
\bibinfo{author}{K.~E. Niemeyer}, \bibinfo{author}{C.-J. Sung},
  \bibinfo{journal}{Combust. Flame} \bibinfo{volume}{158}
  (\bibinfo{year}{2011}) \bibinfo{pages}{1439--1443}.

\bibitem[{Goodwin et~al.(2014)Goodwin, Malaya, Moffat, and Speth}]{cantera2a11}
\bibinfo{author}{D.~Goodwin}, \bibinfo{author}{N.~Malaya},
  \bibinfo{author}{H.~Moffat}, \bibinfo{author}{R.~Speth},
  \bibinfo{title}{Cantera: an object-oriented software toolkit for chemical
  kinetics, thermodynamics, and transport processes. {Version} 2.1a1},
  \bibinfo{howpublished}{\url{https://code.google.com/p/cantera/}},
  \bibinfo{year}{2014}.

\bibitem[{Sj{\"o}berg et~al.(2007)Sj{\"o}berg, Dec, and Hwang}]{sjoberg_2007}
\bibinfo{author}{M.~Sj{\"o}berg}, \bibinfo{author}{J.~E. Dec},
  \bibinfo{author}{W.~Hwang}, \bibinfo{title}{Thermodynamic and chemical
  effects of {EGR} and its constituents on {HCCI} autoignition},
  \bibinfo{type}{Technical Report}, SAE Technical Paper 2007-01-0207,
  \bibinfo{year}{2007}.

\bibitem[{Curran et~al.(1998)Curran, Gaffuri, Pitz, and Westbrook}]{nhept_1}
\bibinfo{author}{H.~J. Curran}, \bibinfo{author}{P.~Gaffuri},
  \bibinfo{author}{W.~J. Pitz}, \bibinfo{author}{C.~K. Westbrook},
  \bibinfo{journal}{Combust. Flame} \bibinfo{volume}{114}
  (\bibinfo{year}{1998}) \bibinfo{pages}{149--177}.

\bibitem[{Curran et~al.(2002)Curran, Gaffuri, Pitz, and Westbrook}]{nhept_2}
\bibinfo{author}{H.~J. Curran}, \bibinfo{author}{P.~Gaffuri},
  \bibinfo{author}{W.~J. Pitz}, \bibinfo{author}{C.~K. Westbrook},
  \bibinfo{journal}{Combust. Flame} \bibinfo{volume}{129}
  (\bibinfo{year}{2002}) \bibinfo{pages}{253--280}.

\bibitem[{Curran et~al.(2004)Curran, Gaffuri, Pitz, and
  Westbrook}]{nhept_website}
\bibinfo{author}{H.~Curran}, \bibinfo{author}{P.~Gaffuri},
  \bibinfo{author}{W.~Pitz}, \bibinfo{author}{C.~Westbrook},
  \bibinfo{title}{n-heptane, detailed mechanism, version 2},
  \bibinfo{howpublished}{\url{https://www-pls.llnl.gov/?url=science_and_technology-chemistry-combustion-n_heptane_version_2}},
  \bibinfo{year}{2004}.

\bibitem[{Peters et~al.(2002)Peters, Paczko, Seiser, and
  Seshadri}]{peters_ignition}
\bibinfo{author}{N.~Peters}, \bibinfo{author}{G.~Paczko},
  \bibinfo{author}{R.~Seiser}, \bibinfo{author}{K.~Seshadri},
  \bibinfo{journal}{Combust. Flame} \bibinfo{volume}{128}
  (\bibinfo{year}{2002}) \bibinfo{pages}{38--59}.

\bibitem[{Smith et~al.(1999)Smith, Golden, Frenklach, Moriarty, Eiteneer,
  Goldenberg, Bowman, Hanson, Song, and Gardiner~Jr}]{gri30}
\bibinfo{author}{G.~P. Smith}, \bibinfo{author}{D.~M. Golden},
  \bibinfo{author}{M.~Frenklach}, \bibinfo{author}{N.~W. Moriarty},
  \bibinfo{author}{B.~Eiteneer}, \bibinfo{author}{M.~Goldenberg},
  \bibinfo{author}{C.~T. Bowman}, \bibinfo{author}{R.~K. Hanson},
  \bibinfo{author}{S.~Song}, \bibinfo{author}{W.~C. Gardiner~Jr},
  \bibinfo{title}{{GRI-Mech} 3.0}, \bibinfo{year}{1999}.

\bibitem[{Sarathy et~al.(2013)Sarathy, Park, Weber, Wang, Veloo, Davis, Togbe,
  Westbrook, Park, Dayma, Luo, Oehlschlaeger, Egolfopoulos, Lu, Pitz, Sung, and
  Dagaut}]{iso-pentanol_mech}
\bibinfo{author}{S.~M. Sarathy}, \bibinfo{author}{S.~Park},
  \bibinfo{author}{B.~W. Weber}, \bibinfo{author}{W.~Wang},
  \bibinfo{author}{P.~S. Veloo}, \bibinfo{author}{A.~C. Davis},
  \bibinfo{author}{C.~Togbe}, \bibinfo{author}{C.~K. Westbrook},
  \bibinfo{author}{O.~Park}, \bibinfo{author}{G.~Dayma},
  \bibinfo{author}{Z.~Luo}, \bibinfo{author}{M.~A. Oehlschlaeger},
  \bibinfo{author}{F.~N. Egolfopoulos}, \bibinfo{author}{T.~Lu},
  \bibinfo{author}{W.~J. Pitz}, \bibinfo{author}{C.-J. Sung},
  \bibinfo{author}{P.~Dagaut}, \bibinfo{journal}{Combustion and Flame}
  \bibinfo{volume}{160} (\bibinfo{year}{2013}) \bibinfo{pages}{2712 -- 2728}.

\bibitem[{Yang et~al.(2010)Yang, Dec, Dronniou, and Simmons}]{isopentanol_hcci}
\bibinfo{author}{Y.~Yang}, \bibinfo{author}{J.~Dec},
  \bibinfo{author}{N.~Dronniou}, \bibinfo{author}{B.~Simmons},
  \bibinfo{journal}{{SAE} Int. J. Fuels Lubr.} \bibinfo{volume}{3}
  (\bibinfo{year}{2010}) \bibinfo{pages}{725--741}.

\bibitem[{Tsujimura et~al.(2012)Tsujimura, Pitz, Gillespie, Curran, Weber,
  Zhang, and Sung}]{isopentanol_old_mech}
\bibinfo{author}{T.~Tsujimura}, \bibinfo{author}{W.~J. Pitz},
  \bibinfo{author}{F.~Gillespie}, \bibinfo{author}{H.~J. Curran},
  \bibinfo{author}{B.~W. Weber}, \bibinfo{author}{Y.~Zhang},
  \bibinfo{author}{C.-J. Sung}, \bibinfo{journal}{Energy Fuels}
  \bibinfo{volume}{26} (\bibinfo{year}{2012}) \bibinfo{pages}{4871--4886}.

\end{thebibliography}
\bibliographystyle{elsarticle-num-CNF}

\pagebreak


\clearpage
\begin{table}
\ra{1.2}
\centering
\begin{tabular}[t]{c}
	\begin{tabular}{@{}lllll@{}}
	\toprule
	& $\phi$ & RPM & T\textsubscript{0} (K) & p\textsubscript{0} (kPa) \\
	\midrule
	NH1 & 0.5 & 1500 & 330 & 101.325\\
	NH2 & 1.0 & 1500 & 330 & 101.325\\
	NH3 & 1.2 & 1500 & 330 & 101.325\\
	\bottomrule
	\end{tabular}
\end{tabular}
\caption{Initial conditions for single cell n-heptane engine simulations at HCCI conditions adapted from Liang et al.~\cite{liang_dac}. Simulations begin at \SI{150}{\degree} before top dead center (BTDC) and end at \SI{150}{\degree} after top dead center (ATDC) to approximate the engine cycle between intake valve closing and exhaust valve opening.}
\label{tab:nhept_hcci_conditions}
\end{table}

\clearpage
\begin{table}[p]
\ra{1.2}
\centering
\begin{tabular}[t]{@{}llll@{}}
\toprule
\multirow{2}{*}{Species} & Number of & Number of & Target species \\
 & neighbors & reactions & importance \\
\midrule
\ce{CO} & 63 & 96 & local \\
\ce{CO2} & 25 & 31 & local \\
\ce{HO2} & 337 & 763 & global \\
\ce{n{\hyphen}C7H16} & 36 & 146 & local \\
\ce{O2} & 352 & 624 & global \\
\ce{OH} & 519 & 1051 & global \\
\bottomrule
\end{tabular}
\caption{The number of neighbors, reactions, and target species behavior related to some commonly used DRGEP target species.}
\label{tab:neighbors}
\end{table}

\clearpage
\begin{table}
\centering
\begin{tabular}[t]{@{}lll@{}}
\toprule
Target set & Maximum error (\%) & Average error (\%) \\
\midrule
\liangtargets\ & 22.5 & 2.06\\
\fulltargets\ & 12.5 & 1.52 \\
$\set{\text{RII} = 3}$ & 10.6 & 1.69 \\
$\set{\text{RII} = 4}$ & 5.34 & 1.31 \\
$\set{\text{RII} = 5}$ & 5.45 & 1.16 \\
\bottomrule
\end{tabular}
\caption{Error statistics for constant-volume autoignition studies of n-heptane at \SIlist{5;20}{\atm}, \SIrange{640}{1200}{\kelvin}, and $\phi = \set{0.5, 1, 2}$. The DRGEP threshold was set to $\Edrgep = $ \num{e-4} in all cases.}
\label{tab:nhept-conv-errorstats}
\end{table}

\clearpage
\begin{table}
\ra{1.2}
\centering
\begin{tabular}[t]{@{}lll@{}}
\toprule
Target set & Maximum error (\%) & Average error (\%) \\
\midrule
\liangtargetsipent\ & 12.3 & 4.43\\
\fulltargetsipent\ & 9.51 & 3.40 \\
$\set{\text{RII} = 3}$ & 6.86 & 2.83 \\
$\set{\text{RII} = 4}$ & 6.58 & 2.33 \\
$\set{\text{RII} = 5}$ & 5.31 & 2.00 \\
\bottomrule
\end{tabular}
\caption{Error statistics for constant-volume autoignition studies of isopentanol at \SIlist{5;20}{\atm}, \SIrange{700}{1100}{\kelvin}, and $\phi = \set{0.38, 0.6, 1.0}$. The DRGEP threshold was set to $\Edrgep = $ \num{5e-3} in all cases.}
\label{tab:ipent-conv-errorstats}
\end{table}

\clearpage
\begin{table}
\ra{1.2}
\centering
\begin{tabular}[t]{c}
\begin{subtable}[c]{\textwidth}
	\centering
	\begin{tabular}{@{}llllll@{}}
	\toprule
	& $\phi$ & RPM & T\textsubscript{0} (K) & p\textsubscript{0} (kPa) & EGR \% \\
	\midrule
	IP1 & 0.38 & 1200 & 405 & 101.325 & 0\\
	IP2 & 0.38 & 1200 & 364 & 140.000 & 0\\
	IP3 & 0.38 & 1200 & 333 & 200.000 & 37\\
	\bottomrule
	\end{tabular}
\end{subtable}
\end{tabular}
\caption{Initial conditions for single cell isopentanol engine simulations at HCCI conditions adapted from Yang et al.~\cite{isopentanol_hcci}. Simulations begin at \SI{158}{\degree}~BTDC and end at \SI{122}{\degree}~ATDC.}
\label{tab:ipent_hcci_conditions}
\end{table}

\clearpage

\clearpage

\clearpage

\clearpage


\clearpage


\clearpage

\clearpage

\clearpage

\clearpage

\clearpage

\clearpage

\clearpage

\clearpage

\clearpage
\end{document}